\documentclass[a4paper,11pt]{article}
\pdfoutput=1 

\usepackage[T1]{fontenc} 
\usepackage{amsmath,amssymb,mathtools}
\usepackage{color}
\usepackage{cite}
\usepackage[colorlinks=true,linkcolor=blue, citecolor=red, urlcolor=blue, bookmarks]{hyperref}
\usepackage{graphics}

\usepackage[text={17.1cm,24.6cm},centering]{geometry} 

\usepackage{multirow,makecell}
\usepackage{textcomp}
\usepackage{wasysym}
\numberwithin{equation}{section}

\newcommand{\beq}{\begin{equation}}
\newcommand{\eeq}{\end{equation}}
\newcommand{\bea}{\begin{eqnarray}}
\newcommand{\eea}{\end{eqnarray}}
\def\l{\lambda}

\begin{document} 

\title{\bf Symmetry resolved entanglement in free fermionic systems}

\author{Riccarda Bonsignori$^1$,
 Paola Ruggiero$^1$, 
 and Pasquale Calabrese$^{1,2}$}

\maketitle
 \vspace{-10mm}
\begin{center}
{\it
$^{1}$ International School for Advanced Studies (SISSA) and INFN, Via Bonomea 265, 34136 Trieste, Italy\\\vspace{1mm}
$^{2}$International Centre for Theoretical Physics (ICTP), Strada Costiera 11, 34151 Trieste, Italy
}
\vspace{10mm}
\end{center}




%
%

\begin{abstract}
We consider the symmetry resolved R\'enyi entropies in the one dimensional tight binding model, equivalent to the spin-1/2  XX chain in a magnetic field. 
We exploit the generalised Fisher-Hartwig conjecture to obtain the asymptotic behaviour of the entanglement entropies with a flux charge insertion at leading and subleading orders. 
The $o(1)$ contributions are found to exhibit a rich structure of oscillatory behaviour.
We then use these results to extract the symmetry resolved entanglement, determining exactly all the non-universal constants and logarithmic corrections 
to the scaling that are not accessible to the field theory approach.
We also discuss how our results are generalised to a one-dimensional free fermi gas.
\end{abstract}

\baselineskip 18pt
\thispagestyle{empty}
\newpage


\tableofcontents


\section{Introduction}

Entanglement measures turned out to be fundamental tools for a finer description of the quantum world. 
Nowadays a lot has been understood in the context of low-dimensional many body quantum systems, both from the theoretical 
side \cite{amico-2008,calabrese-2009,eisert-2010,rev-lafl} and, more recently, also experimentally \cite{greiner-15,kauf-16,zoller-18,exp-lukin}.
For bipartite pure states described by a density matrix $\rho$, the most useful measure is surely the \emph{entanglement entropy}, defined as the von Neumann entropy 
of the reduced density matrix (RDM) $\rho_A ={\rm  tr}_{B} \rho$ associated to one of the two subsystems (denoted by $A$ and $B$ respectively), i.e.,
\beq
\label{svN_0}
S_{\rm vN}= - {\rm tr} \rho_A \ln \rho_A.
\eeq
$S_{\rm vN}$ is the limit for $n \to 1$ of a larger family of entropies, known as R\'enyi entropies
\beq
\label{sn_0}
S_n = \frac{1}{1-n} \ln {\rm tr} \rho_A^n.
\eeq
Within quantum field theory (QFT) both Eqs. \eqref{svN_0} and \eqref{sn_0} are usually obtained from the integer moments $Z_n= {\rm tr} \rho_A^n$ of $\rho_A$, which are in turn easily expressed in the path integral formalism as partition function on suitable $n$-sheeted Riemann surfaces \cite{cc-04,cc-09}. 
This approach, when applied to critical systems, whose low energy physics is described by $(1+1)$ dimensional conformal field theory (CFT), leads to the famous scaling results \cite{hlw-94,cc-04,cc-09,vlrk-03}
\beq
S_{\rm vN} = \frac{c}{3} \ln \ell, \qquad S_n = \frac{c}{6} \frac{(n+1)}{n} \ln \ell,
\eeq
for a subsystem $A$ made of a single interval of length $\ell$ embedded in an infinite one-dimensional system (and similarly for finite systems, systems at finite temperature, 
and other situations: it is sufficient to replace $\ell$ with the relevant length in the considered regime, see e.g. \cite{cc-09}).

However, a recent experiment, in the context of disordered systems \cite{exp-lukin}, showed that it is also important to understand the ``internal symmetry structure'' of the 
entanglement as well. 
In particular, looking at systems possessing an internal global symmetry, entanglement turned out to have two different contributions, 
dubbed \emph{configurational} and \emph{number} or {\it fluctuation} entanglement \cite{exp-lukin}.
These two contributions account for the entanglement within symmetry sectors and fluctuations thereof, respectively (see below for a precise definition).

At the same time, a new theoretical framework has been developed to address the problem of extracting the \emph{symmetry resolved} contributions for
different entanglement measures \cite{GS,GS-neg,equi-sierra,fg-19}. 
Indeed, these contributions have been related to the moments of $\rho_A$ where twisted boundary conditions are imposed along the cuts of the Riemann surface: 
we will refer to them as \emph{charged} moments.
As we are going to see more in details, the twist can be implemented geometrically within field theory via threading an appropriate Aharonov-Bohm flux through the multisheeted Riemann surface \cite{GS}.
The relation between twisted boundary conditions and flux insertion was actually previously explored, for example, in the context of free field theories \cite{CFH,d-16,ch-rev}. 
Moreover, similar quantities have also been introduced in the holographic setting \cite{matsuura,cnn-16} and  in the study of entanglement 
in mixed states \cite{GS-neg,ssr-17,shapourian-19}.

If on one hand the field theory approach is very powerful and versatile in order to provide the scaling limit of both the charged and symmetry resolved entanglement entropies, 
on the other, it does not give access to non-universal model-dependent pieces which are also very important to accurately characterise critical systems.
As we are going to see, for the special case of free fermions, we can go beyond the field theory  results. 
Indeed, we can rely on the \emph{(generalised) Fisher-Hartwig conjecture} \cite{basor, JK, ce-10} to compute systematic expansions of these entropies 
which reproduce the field theory results and provide exact expressions for the non-universal terms. 
In fact, this method, which has already been explored for the standard entanglement and R\'enyi entropy  \cite{JK, ce-10}, can be simply generalised to the same quantities with 
a further flux insertion and therefore to their symmetry resolved analogue.

The paper is organised  as follows. In Section \ref{sec:known} we carefully define all the quantities we will be dealing with and give an overview of the field theory results. 
Sections \ref{sec:flux} and \ref{sec:symm_resolved} are the core of this paper where we derive results for free fermions on a lattice for the charged and 
the symmetry resolved entanglement entropies, respectively. 
In Section \ref{sec:gas} we show how all results derived for the lattice model can be directly adapted to a free Fermi gas.
We conclude in Section \ref{sec:end} with some remarks and discussions. 
Some details of the calculations can be found in the appendix.


\section{Symmetry resolution and flux insertion}
\label{sec:known}

Let us consider a many body quantum system with an internal $U(1)$ symmetry.
Let $\rho$ be the density matrix in a given (pure) state and $Q$ the operator generating such symmetry.
If the system is in a given representation of the charge $Q$, i.e., in an eigenstate of $Q$ corresponding to a definite eigenvalue, then $[\rho,Q]=0$. 

We will be interested in a bipartition of the total system into two complementary spatial subsystems $A$ and $B$, with $\rho_A= {\rm tr}_B \rho$ being the reduced density matrix 
of the subsystem $A$. 
Usually the operator $Q$ splits in the sum $Q= Q_A+Q_B$, meaning that $Q$ comes from local degrees of freedom within the two subsystems.
Consequently, by taking the trace over $B$ of $[\rho,Q]=0$, we find that $[\rho_A, Q_A]=0$. 
This implies that $\rho_A$ acquires a block diagonal form, in which each block corresponds to a different charge sector with a definite eigenvalue $q$ of $Q_A$, i.e.,
\beq
\label{direct_sum}
\rho_A = \oplus_{q}  \Pi_q \rho_A =  \oplus_{q} \left[ p (q) \rho_A (q) \right] ,
\eeq
where $\Pi_q$ is the projector on  eigenspace of fixed value of $q $ in the spectrum of $Q_A$. 
In the last equality we factorised $p(q)= {\rm tr} (\Pi_q \rho_A)$, the probability of finding $q$ as the outcome of a  measurement of $Q_A$. 
Note that in this way the density matrices $\rho_A (q)$ of different blocks are normalised as ${\rm tr}\rho_A (q)=1$.

Our goal is to understand how the entanglement is distributed in the different charge sectors. 
Focusing on the von Neumann entanglement entropy as a prototypical example, Eq.~\eqref{direct_sum} implies the following decomposition
\beq
\label{decomposition}
S_{\rm vN} = \sum_q p(q) S_{\rm vN} (q) - \sum_q p(q) \ln p(q) \equiv S^c + S^f,
\eeq
where we defined the \emph{symmetry resolved entanglement entropy} as the one associated to $\rho_A (q)$ in \eqref{direct_sum}
\beq
\label{vN_q}
S_{\rm vN} (q) \equiv - {\rm tr} \left[ \rho_A (q) \ln \rho_A (q) \right].
\eeq
The two different contributions in \eqref{decomposition} are the \emph{configurational} entanglement entropy, $S^c\equiv \sum_q p(q) S_{\rm vN} (q)$ \cite{wv-03, exp-lukin}, 
measuring the total entropy due to each charge sector (weighted with their probability) 
and the \emph{fluctuation} entanglement entropy $S^f=- \sum_q p(q) \ln p(q)$ \cite{exp-lukin}, which instead takes into account the entropy due to the fluctuations of the  value of the charge
within the subsystem $A$.

Similarly, one defines also \emph{symmetry resolved R\'enyi entropies} as
\beq
\label{renyi_q}
S_n (q) \equiv \frac{1}{1-n} \ln {\rm tr} \left[ \rho_A (q) \right]^n.
\eeq
%


In general evaluating such symmetry resolved quantities is a highly non-trivial problem, mainly due to the non local nature of the projector $\Pi_q$.
As mentioned in the introduction, recently, this problem has been understood from a different perspective in \cite{GS,equi-sierra}.
This new approach works as follows. Let us first define the (unnormalised) quantity 
\beq
\label{Z_q}
\mathcal{Z}_n (q) \equiv {\rm tr} \left( \Pi_q \rho_A^n \right),
\eeq
which is related to the entanglement and R\'enyi entropies in \eqref{vN_q} and \eqref{renyi_q} (respectively) through
\beq
\label{replica}
S_{n} (q) = \frac{1}{1-n} \ln \left[\frac{\mathcal{Z}_n(q)}{\mathcal{Z}_1 (q)^n} \right] \qquad 
S_{\rm vN} (q)= - \partial_n  \left[\frac{\mathcal{Z}_n(q)}{\mathcal{Z}_1 (q)^n} \right]_{n=1}.
\eeq
Also the probability $p(q)$ is read off ${\cal Z}_n$ as
\beq
\label{probability}
p(q) = \mathcal{Z}_1 (q).
\eeq
The key observation of Refs. \cite{GS,equi-sierra} is that \eqref{Z_q} is given by the following Fourier transform 
\beq
\label{fourier}
\mathcal{Z}_n (q) = \int_{-\pi}^{\pi} \frac{d \alpha}{2 \pi} e^{-i q \alpha} \, Z_n (\alpha), \qquad  Z_n (\alpha) \equiv {\rm tr}  \left(  \rho_A^n e^{i  Q_A \alpha} \right), 
\eeq
where $Z_n (\alpha)$ are the charged moments mentioned in the introduction. Note that $Z_n (0)= {\rm tr} \rho_A^n$.
Therefore, we can access the symmetry resolved entanglement entropy by studying $Z_n (\alpha)$ (which, as explained below, are much easier to compute) 
and after Fourier transforming. 

\subsection{Replica method and results from QFT}

In Ref. \cite{GS} a geometric approach in the framework of the replica trick has been introduced and it is applicable to generic (1+1)-dimensional QFT. 
The main idea is to insert an appropriate conjugate Aharonov-Bohm flux through a multi-sheeted Riemann surface $\mathcal{R}_n$, 
such that the total phase accumulated by the field upon going through the entire surface is $\alpha$.
The result is that $Z_n (\alpha)$ is  the partition function on such modified surface.

In QFT language, the insertion of the flux corresponds to a twisted boundary condition, which, as usually done in this context, can be implemented by the action of 
a local operator, acting at the boundary of the subsystem $A$. 
This operator is a modified \emph{twist field} $\mathcal{T}_{n, \alpha} $ whose action, in operator formalism, is defined by \cite{GS}
\beq
\mathcal{T}_{n, \alpha} (x, \tau)  \phi_i (x', \tau) = 
\begin{cases}
\phi_{i+1} (x', \tau) e^{i \alpha \delta_{i, j}} \mathcal{T}_{n, \alpha} (x, \tau) \qquad (x< x'), \\
\phi_{i} (x', \tau) \mathcal{T}_{n, \alpha} (x, \tau) \quad \qquad \qquad \textrm{otherwise}.
\end{cases}
\eeq
In this way one can further reformulate the problem in terms of a correlation function of twist fields \cite{CCAD08}. 
In the simplest case of the subsystem consisting of a single interval $A= [0, \ell]$
\beq
Z_n (\alpha) = \langle \mathcal{T}_{n, \alpha} (\ell, 0) \tilde{\mathcal{T}}_{n, \alpha} (0, 0) \rangle.
\eeq
where $\tilde{\mathcal{T}}$ is the antitwist field. If we now specialise to (1+1) dimensional CFT, $\mathcal{T}_{n, \alpha}$ and $\tilde{\mathcal{T}}_{n, \alpha}$ behave as primary operators with conformal 
dimension given by \cite{GS}
\beq
h_{n, \alpha} = h_n +\frac{h_{\alpha}}{n} , \qquad h_n = \frac{c}{24}\left( n- \frac{1}{n} \right),
\eeq
meaning that the phase shift is implemented by a composite twist field that can be written as $\mathcal{T}_{n, \alpha}= \mathcal{T}_n \cdot \mathcal{V}_{\alpha}$. 
This immediately implies
\beq
\label{Z_n_alpha}
Z_n (\alpha) = c_{n,\alpha} \ell^{- \frac{c}{6} \left( n - \frac{1}{n} \right) - 2 \frac{h_{\alpha} + \bar{h}_{\alpha} }{n}} ,
\eeq
where $c$ is the central charge of the CFT and $c_{n,\alpha}$ the unknown non-universal normalisation of the composite twist-field. 

The focus of Ref.  \cite{GS} was a free  boson compactified on a circle of radius $R$, i.e., a Luttinger liquid with Luttinger parameter $K$. 
In this case the operator $\mathcal{V}_{\alpha}$ implementing the twisted boundary conditions is a vertex operator with (holomorphic and antiholomorphic) 
scaling dimensions  
\beq
h_{\alpha} = \bar{h}_{\alpha}= \frac{1}{2} \left( \frac{\alpha}{2 \pi} \right)^2 K.
\eeq
From Eq. \eqref{Z_n_alpha}, the symmetry resolved moments are found by taking the Fourier transform as in Eq.~\eqref{fourier}. 
At leading order for large $\ell$, this reads \cite{GS}
\beq
\label{Znq_FT}
\mathcal{Z}_n (q) \simeq \ell^{- \frac{c}{6} \left( n - \frac{1}{n} \right)} \sqrt{\frac{n \pi}{2 K \ln \ell}} e^{\frac{n \pi^2 (q- \langle Q_A \rangle)^2 }{2 K \ln \ell}}.
\eeq
Notice that we set a posteriori the average number of the charge in the subsystem $\langle Q_A\rangle$, since it is a non-universal quantity, not encoded in the CFT.  
For a given microscopical model, its origin can be easily traced back, e.g. as a phase shift in the bosonisation rule. 

Through Eq.~\eqref{replica}, this leads to the following result at leading order for the R\'enyi and the von Neumann entropy 
\beq
\label{result_CFT}
S_n (q) = S_n - \frac12 \ln \left( \frac{2 K}{\pi} \ln \ell \right)  + O(\ell^0), \qquad S_{\rm vN} (q) = S_{\rm vN}- \frac12 \ln \left( \frac{2 K}{\pi} \ln \ell \right)  + O(\ell^0).\\
\eeq
This result has been dubbed \emph{equipartition of entanglement} \cite{equi-sierra}: at leading order the entanglement is the same in the different charge sectors, 
just the probability $p(q)$ of being in a given sector varies.


\section{Free fermions on a lattice: flux insertion and charged entropies}
\label{sec:flux}

Eq.~\eqref{result_CFT} provides the leading symmetry resolved entanglement entropies of all microscopical models with a $U(1)$ symmetry, that, at low energy, 
are described by a CFT.
Indeed the results in Eq.~\eqref{result_CFT} have been tested numerically both for free fermions \cite{GS,equi-sierra} and in interacting spin chains \cite{equi-sierra,lr-14}.   
In this Section we are going to provide an analytic derivation for the special case of a chain of free fermions, whose scaling limit is indeed 
described by a free compact boson with $K=1$. 
Our analysis will also provide the exact value of the non-universal constants, as well as the corrections to \eqref{result_CFT} for this specific model.


We consider the tight binding model in one dimension with hamiltonian 
\beq
\label{ham}
H = - \sum_{i = -\infty}^{\infty} \left[ c_i^{\dagger} c_{i+1} + c_{i+1}^{\dagger} c_i - 2 h \left(c_i^{\dagger} c_i -\frac{1}{2} \right) \right],
\eeq
where $c_i$ are free fermionic spinless degrees of freedom, satisfying the anticommutation relations $\{ c_i, c_j^{\dagger}\} = \delta_{ij}$ and $h$ is the chemical potential.
$H$ is diagonal in momentum space and its ground state is a Fermi sea with  Fermi momentum $k_F = \arccos |h|$.
As it is clear from \eqref{ham}, the particle number $Q = \sum_i c_i^{\dagger} c_i$ is a conserved $U(1)$ charge of the model.
It is also local and $Q=Q_A+Q_B$ for any spatial bipartition of the chain. 
By Jordan Wigner transformation, Eq. \eqref{ham} is mapped to the XX spin chain in a magnetic field $h$ and the charge $Q$ becomes the spin in the $z$ direction.

In the following we will be interested in the bipartition where the subsystem $A$ is given by $\ell$ contiguous lattice sites. 
The corresponding RDM of the subsystem can then be written as \cite{peschel2001, peschel2003,pe-09}
\beq
\label{RDM}
\rho_A = \det C_A \exp \Big(\sum_{i, j} \left[ \ln (C_A^{-1} - 1) \right]_{ij} c_i^{\dagger} c_j \Big),
\eeq
where the $\ell \times \ell$ matrix $(C_A)_{ij} \equiv \langle c_i^{\dagger} c_j \rangle$ is the \emph{correlation matrix} restricted to the subsystem $A$, 
that for the ground-state of an infinite chain has elements
\beq
\label{cm_lattice}
(C_A)_{i, j} = \frac{\sin k_F (i-j) }{\pi (i-j)}, \qquad i,j\in A.
\eeq
Note that $C_A$ is a Toeplitz matrix, meaning that its entries only depend on the difference $(i-j)$: this is a key point for what follows.

If one write the eigenvalues of the matrix $C_A$ as $(1+ \nu_k)/2 $ (with $k \in [1, \ell]$), then simple algebra leads to the 
moments of $\rho_A$ as
\beq
\label{moments_0}
{\rm tr} \rho_A^n = \prod_{i=1}^{\ell}  \left[ \left( \frac{1+\nu_i}{2}\right)^n  + \left( \frac{1- \nu_i}{2}\right)^n \right],
\eeq
and, equivalently, the R\'enyi entropies read
\beq
\label{renyi_lattice0}
S_{n} = \sum_{i=1}^{\ell} e_n (\nu_i), \qquad e_n (x) \equiv \frac{1}{1-n} \ln \left[ \left( \frac{1+x}{2}\right)^n  + \left( \frac{1- x}{2}\right)^n \right] .
\eeq

It has been first noticed in Ref. \cite{GS} that the $\alpha$-dependent moments $Z_n (\alpha)$, defined in \eqref{fourier}, can be also easily written in terms of the eigenvalues 
of the correlation matrix with a simple modification of the above formulas, i.e.,
\beq
\label{znalpha_nu}
Z_n (\alpha) =  \prod_{i=1}^{\ell} \left[ \left( \frac{1+\nu_i}{2}\right)^n e^{i \alpha}  + \left( \frac{1- \nu_i}{2}\right)^n \right]\,.
\eeq
The interpretation of this equation is straightforward: each particle carries a weight $e^{i \alpha}$ while the holes carry weight $1$.
Eq.~\eqref{znalpha_nu}  provides a very simple method for the numerical computation of $Z_n (\alpha)$.
Not only: as we are going to discuss next, it is also the right starting point to get the asymptotic analytic expressions of $Z_n (\alpha)$.

Before embarking into the study of $Z_n (\alpha)$ a quick recap of its properties and limits is necessary, also to provide useful consistency checks 
for our calculations.
First, for $n=1$
\beq
Z_1 (\alpha) \equiv {\rm tr }\rho_A e^{i Q_A\alpha} =  \prod_{i=1}^{\ell} \left[ \left( \frac{1+\nu_i}{2}\right) e^{i \alpha}  + \left( \frac{1- \nu_i}{2}\right) \right]\,,
\label{Z1}
\eeq
is the moment-generating function of $Q_A$. 
This quantity has been already studied in the literature \cite{sfr-11a,sfr-11b,cmv-12,si-13,clm-15} also because of its relation with the entanglement entropy itself. 
We will see that it is simply related to another quantity usually introduced in this context ($D_\ell(\lambda)$ of the next subsection). 
The first moment is just the average number of particle in $A$, i.e., $\langle Q_A\rangle= \ell k_F/\pi$.
Hence $Z_1(\alpha)=1+ i\alpha \langle Q_A\rangle+O(\alpha^2)$. 
At half-filling $Z_1(\alpha)$ further simplifies as a consequence of the fact that, by particle-hole symmetry, for each 
 $\nu_{i}$ there is a $\nu_j$ such that $(1- \nu_{i})= (1+ \nu_j)$.  Thus we have 
\begin{equation}
\label{s1_alpha2}
Z_1 (\alpha) =
e^{i \alpha \ell /2} \prod_{i}  \left[  \left( \frac{1 + \nu_{i}}{2} \right) e^{i \alpha/2} +  \left( \frac{1 - \nu_{i}}{2} \right) e^{-i \alpha/2} \right]= e^{i \alpha \ell /2} g(\alpha)\,,
\end{equation}
with $g(\alpha)$ real and even.
For general filling instead the odd cumulants are non-vanishing (the odd derivatives of $\ln Z_1(\alpha)$ are non zero) 
and $Z_1(\alpha)$ has no particular parity or reality properties. 
Indeed, Eq. \eqref{s1_alpha2} remains true for generic $n$
\begin{equation}
\label{sn_alpha2}
Z_n (\alpha) =
e^{i \alpha \ell /2} \prod_{i}  \left[  \left( \frac{1 + \nu_{i}}{2} \right)^n e^{i \alpha/2} +  \left( \frac{1 - \nu_{i}}{2} \right)^n e^{-i \alpha/2} \right]= e^{i \alpha \ell /2} g_n(\alpha)\,,
\end{equation}
with $g_n(\alpha)$ real and even.
This symmetry of $Z_n (\alpha)$ at half-filling represents a cross check of our numerical and analytic calculations. 
Again, away from half-filling,  $\ln Z_n (\alpha)$ has all non-zero derivatives. 

We notice that by Fourier transforming \eqref{znalpha_nu} one easily gets 
\begin{equation}
\label{znq_nu}
\mathcal{Z}_n (q) =\sum_{\mathcal{S}_q} \prod_{i \in \mathcal{S}_q} \left( \frac{1+ \nu_i}{2} \right)^n \prod_{j \in \bar{\mathcal{S}_q}}  \left( \frac{1 -  \nu_j}{2} \right)^n,
\end{equation}
where the sum is over all subset $\mathcal{S}_q$ of $\mathcal{S} = {1, \cdots, \ell}$ of cardinality $q$ and $\bar{\mathcal{S}_q}$ denotes the complementary subset.
Unfortunately Eq. \eqref{znq_nu} is not a very convenient way to get $\mathcal{Z}_n (q)$, since one has to sum over $\ell!/((\ell-q)!q!)$ terms and this is 
impossible already for moderate values of $\ell$. The most convenient way to extract $\mathcal{Z}_n (q)$ is by direct Fourier transform of $Z_n (\alpha) $.

\subsection{Charged entropies via the generalised Fisher-Hartwig conjecture}

The method that we employ takes advantage of the Toeplitz nature of the correlation matrix that can be handled with the 
\emph{(generalised) Fisher-Hartwig conjecture} providing the asymptotics of determinant of Toeplitz matrices.
This technique has been used already in the context of entanglement in free lattice models to derive the leading term and the corrections to 
entanglement entropies \cite{JK, ce-10,km-05,afc-09,fc-11,aef-14,aef-14b,aef-15}.
We are going to show that the same technology applies also to the $\alpha$-dependent moments $Z_n (\alpha)$ and therefore, as a consequence of the discussion above, 
to their symmetry resolved equivalent $\mathcal{Z}_n (q)$.

 
The starting point of our analysis is to rewrite the logarithm of Eq.~\eqref{znalpha_nu}
\beq
\label{zn_sumfn}
\ln Z_n (\alpha) = \sum_{i =1}^{\ell} f_n (\nu_i, \alpha), \quad f_n (x, \alpha) =  \ln \left[ \left( \frac{1+x}{2}\right)^n e^{i \alpha} + \left( \frac{1- x}{2}\right)^n \right] ,
\eeq
as a contour integral
\beq
\label{fn_lattice}
\ln Z_n (\alpha) = \frac{1}{2\pi i}\oint d \lambda \, f_n(\lambda, \alpha)  \frac{d \ln D_{\ell}(\lambda)}{d \lambda}, 
\eeq
where the contour of integration encircles the segment $[-1, 1]$. Here we defined the characteristic polynomial of $C_A$ as the determinant
\beq
\label{det}
D_{\ell} = \det \big[
(\lambda + 1) \mathbb{I}_A - 2 C_A\big],
\eeq
where $\mathbb{I}_A$ is the identity matrix restricted to $A$.
In the basis that diagonalises $C_A$, such determinant simply becomes $D_{\ell}= \prod_i (\lambda - \nu_i)$ and therefore, by residue theorem, Eq. \eqref{fn_lattice} 
is the same as \eqref{zn_sumfn}.
Notice that $D_\ell(\lambda)$ is related to the generating function $Z_1(\alpha)$ as
\beq
Z_1(\alpha)= \Big(\frac{1-e^{i\alpha}}{2}\Big)^\ell D_\ell \Big(i \cot \frac\alpha2 -\l\Big).
\eeq

In Refs. \cite{JK, ce-10} it has been exploited the fact that the matrix $(\lambda + 1) \mathbb{I}_A - 2 C_A$ has a Toeplitz form.  
Therefore the asymptotics for large $\ell$ of the determinant $D_{\ell}$ in \eqref{det} is obtained by means  of the generalised Fisher-Hartwig conjecture.
The interested reader can find the derivation in Ref. \cite{ce-10}, we  just report here the final result which is \cite{ce-10}
\beq
\label{gen_FHconj}
D_{\ell}(\lambda)\simeq (\lambda+1)^{\ell}\left(\frac{\lambda+1}{\lambda-1}\right)^{-\frac{k_F \ell}{\pi}}\sum_{m \in \mathbb{Z}}L_k^{-2(m+\beta_{\lambda})^2}e^{-2ik_F m \ell}\left[ G(m+1+\beta_{\lambda})G(1-m-\beta_{\lambda}) \right]^2,
\eeq
where $G (\cdot)$ is the Barnes $G$-function, $L_k = 2 \ell |\sin k_F|$ and 
\beq
\beta_{\lambda} = \frac{1}{2 \pi i} \ln \left[ \frac{\lambda+1}{\lambda-1} \right],\qquad {\rm with}\quad
\frac{d \beta_\lambda}{d \lambda}=\frac{1}{\pi i}\frac1{1-\lambda^2}.
\eeq

For the moments ${\rm tr}\rho_A^n$, i.e., $ Z_n (\alpha=0)$ in \eqref{zn_sumfn}, the leading term in the sum for $D_\ell$ in Eq. \eqref{gen_FHconj}
is the one with $m=0$, first evaluated in \cite{JK}. 
The next to leading contributions  come from the terms with $m=\pm 1$  (at the same order) as shown in \cite{ce-10}.
The situation for $\alpha\neq 0$ is slightly more complicated. 
For $-\pi<\alpha<\pi$ the leading term is always the one with $m=0$.
Since $Z_n (\alpha)$ is periodic in $\alpha$ with period $2\pi$ we will restrict ourselves to $\alpha\in [-\pi,\pi]$, having in mind that, if required, the function 
can be extended outside of this interval by periodicity. 
Concerning the subleading contributions, the terms with $m=\pm1$ have different power laws, but one of them is always dominating, as we shall see. 
Anyhow, for values of $\alpha$ close to $\pm\pi$, also next-to-next leading terms should be taken into account in order to get reasonable results
for moderately large values of $\ell$. 
In the following we first compute the leading term and then we move to the calculations of the corrections. 
 

\subsubsection{Leading term $(m=0)$}

For $\alpha\in[-\pi,\pi]$, the leading behaviour of Eq.~\eqref{fn_lattice} is given by term with $m=0$ in \eqref{gen_FHconj}, i.e.,
\beq
\label{FH0}
D_{\ell}^{(0)}(\lambda)\equiv (\lambda+1)^{\ell}\left(\frac{\lambda+1}{\lambda-1}\right)^{-\frac{k_F \ell}{\pi}}
L_k^{-2\beta_{\lambda}^2}\left[ G(1+\beta_{\lambda})G(1-\beta_{\lambda})\right]^2,
\eeq
so that the integral \eqref{fn_lattice} is $\ln Z_n(\alpha)=\ln Z_n^{(0)}(\alpha)+o(\ell^0)$ with 
\begin{align}
\label{tn_alpha1}
\ln Z_n^{(0)}(\alpha)&=  \frac{1}{2\pi i}\oint d \lambda \, f_n(\lambda, \alpha)  \frac{d \ln D^{(0)}_{\ell}(\lambda)}{d \lambda}=
a_0 \ell + a_1\ln L_k +a_2,
\end{align}
where 
\begin{align}
a_0&= \frac{1}{2\pi i}\oint d\lambda f_n(\lambda,\alpha)\left(  \frac{1-k_F/\pi}{1+\lambda}-\frac{k_F/\pi}{1-\lambda}\right), \\
a_1&= \frac{1}{2\pi i}\oint d\lambda f_n(\lambda,\alpha)\frac{d(-2  \beta_{\lambda}^2)}{d\lambda}= \frac{2}{\pi^2} \oint d\lambda f_n(\lambda,\alpha) \frac{\beta_\lambda}{1-\lambda^2}
,\\ 
a_2&=\frac{1}{\pi i}\oint  d\lambda f_n(\lambda,\alpha) \frac{d \ln [G(1+\beta_{\lambda})G(1-\beta_{\lambda})]}{d\lambda},
\end{align}
are respectively the linear, the logarithmic and the constant term (in $\ell$) coming from $\ln D^{(0)}_\ell$ in Eq. \eqref{FH0}. 
These three integrals are explicitly calculated in Appendix \ref{details}
%
%
with final result 
\begin{align}
\label{tn_0}
\ln Z_n^{(0)}(\alpha)&=  i \alpha \frac{k_F \ell}{\pi}-
\left[ \frac{1}{6}\left( n-\frac{1}{n} \right) + \frac{2}{n}\left( \frac{\alpha}{2 \pi} \right)^2\right] \ln L_k +   \Upsilon{(n,\alpha)} ,
\end{align}
where 
\begin{equation}
\label{upsilon}
 \Upsilon{(n,\alpha)}= {n i}\int_{-\infty}^\infty  dw [\tanh(\pi w)-\tanh (\pi n w+i\alpha/2)]  \ln \frac{\Gamma(\frac12 +iw)}{\Gamma(\frac12 -iw)} \,,
\eeq
in analogy with the definition $\Upsilon (n)$ in \cite{JK}, which is recovered when $\alpha=0$. 
We stress that $\Upsilon{(n,\alpha)}$ is real for $\alpha$ real, even if not apparent from the formula. 
For future reference it is useful to write $\Upsilon{(n,\alpha)}$ as
\beq
\Upsilon (n, \alpha) = \Upsilon(n) + \gamma_2 (n) \alpha^2 + \epsilon (n, \alpha), \qquad \epsilon (n, \alpha)= O (\alpha^4),
\label{Upsexp}
\eeq

\begin{figure}[tbp]
  \centering
  \includegraphics[width=0.99\textwidth]{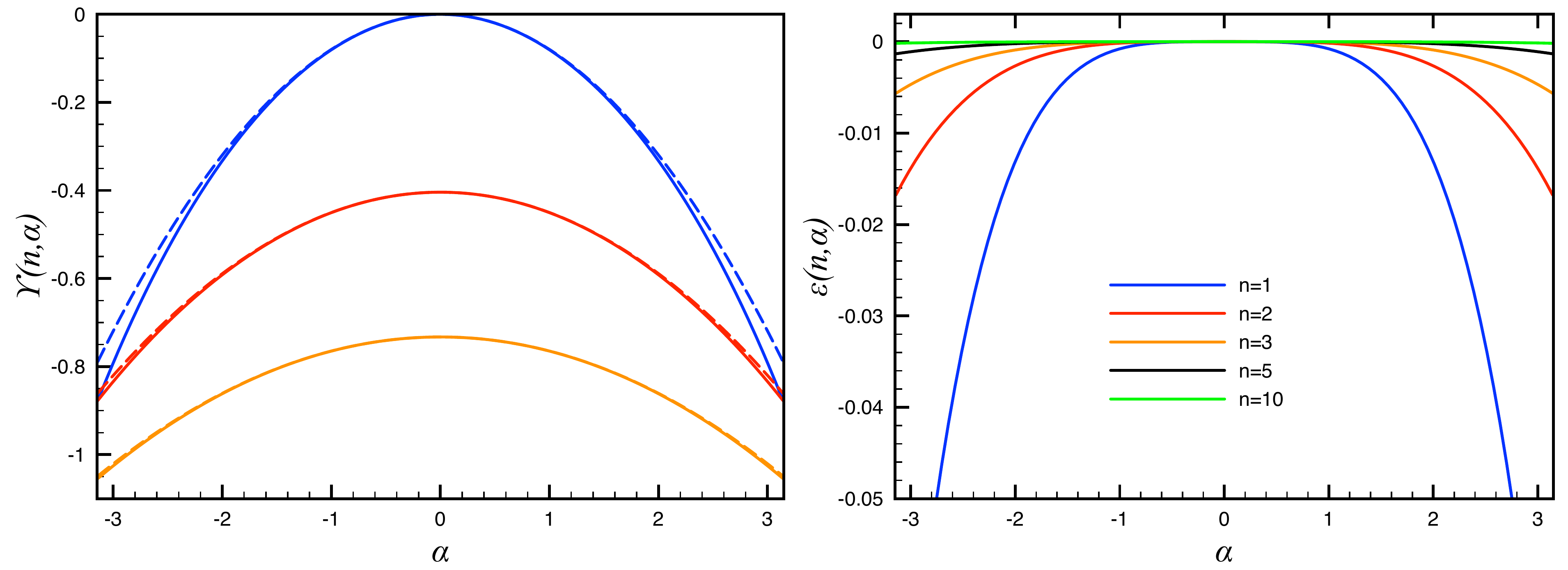}\\
  \caption{$\Upsilon(n,\alpha)$ in Eq. \eqref{upsilon} as a function of $\alpha$ for $n=1,2,3$ (top to bottom in the left panel). 
  The exact forms (full lines) are compared with the quadratic approximation $\Upsilon(n)+\gamma_2(n) \alpha^2$ (dashed lines)
  showing that, although very close, they are definitively different. 
  To highlight this similarity we plot in the right panel the difference 
  $ \epsilon(n,\alpha)$, cf. Eq \eqref{Upsexp} which is tiny, but non zero.
  }
  \label{Fig:ups}
\end{figure}

Eq. (3.18) contains several pieces of information. 
The linear term is just the mean number of particles in $A$, $\langle Q_A\rangle= k_F \ell/\pi$, as expected. 
Anyhow, this is the only term with an imaginary part up to order $O(1)$. 
We know that this is {\it exactly} true at half-filling ($k_F=\pi/2$), cf. Eq. \eqref{sn_alpha2}. For generic filling, it is not true in general and
we will observe in numerics tiny deviations at small $\ell$ in the imaginary part of $\ln Z_n(\alpha)$. 
The  term $\propto \ln L_k$ provides the dimension of the modified twist field which comes out from the field theory calculation: 
the result agrees with the one found by CFT methods in \eqref{Z_n_alpha} when specialised to a compact boson with $K=1$.  
It was important to test analytically this result that was already checked numerically in \cite{GS}.
The constant term in Eq. \eqref{tn_0} is probably the most interesting one, first because it is a result that was not known by other means 
(being non-universal cannot be fixed by field theory), and second because it provides few physical consequences. 
It is real and even in $\alpha$, a property that was guaranteed only at half filling. 
It is independent from $k_F$, as its limit for $\alpha=0$ \cite{JK}.   
Finally, it is very close to a parabola, but all the even terms in the series expansion are non zero, although $\epsilon(n,\alpha)$, cf. Eq. \eqref{Upsexp}, is very small. 
In Figure \ref{Fig:ups} we report $\Upsilon{(n,\alpha)}$ as function of $\alpha$ for some $n$ and compare it with the quadratic approximation 
$\Upsilon{(n)}+\gamma_2(n) \alpha^2$. The closeness of the two curves shows that the quadratic approximation 
will be enough for most of the applications, as we shall explicitly show.
The accuracy of the quadratic approximation is also evident from the plot of $\epsilon(n,\alpha)$ in the right panel of Figure \ref{Fig:ups}.
On passing, this precision of the quadratic approximation of $\Upsilon{(n,\alpha)}$ explains, a posteriori, the quality of the symmetry resolved spectrum 
obtained in Ref. \cite{GS} exploiting the method of Stieltjes transform \cite{cl-08} which implicitly assumes this approximation.

In Figure \ref{Fig:snal} we report the numerical data for R\'enyi entropies with the insertion of a flux $\alpha$ for several values of $n$ and $\alpha$ and 
with fillings $k_F=\pi/2$ (left) and $k_F=\pi/3$ (right). 
The theoretical prediction for the leading scaling in Eq. \eqref{tn_0} is also reported for comparison. 
It is evident that the analytical result correctly describes the asymptotic data, but large and oscillating corrections to the scaling are present. 
The amplitude of these oscillations increase with $n$ and with $\alpha$.
This peculiar $n$ dependence was already known at $\alpha=0$ \cite{ce-10,ccen-10,cc-10,ccp-10,ot-15}. 
In the following subsection we will explicitly consider these oscillations and work out their analytical description.

\begin{figure}[tbp]
  \centering
  \includegraphics[width=0.99\textwidth]{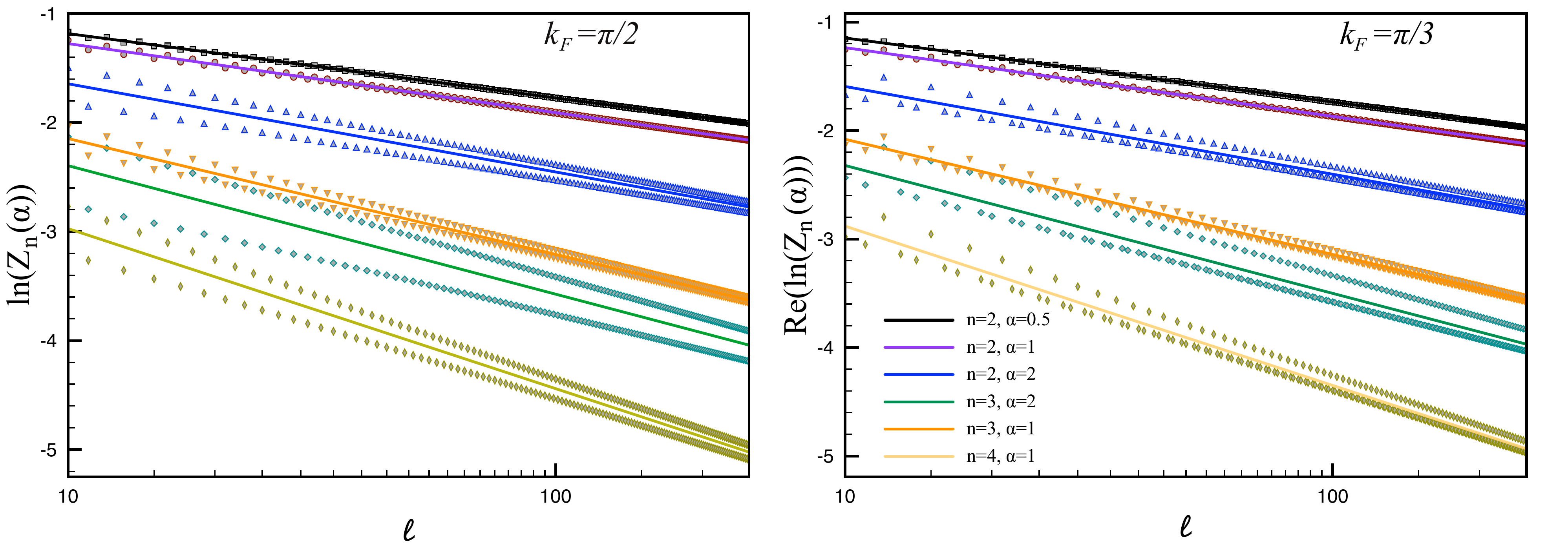}\\
  \caption{Leading scaling behaviour of the charged R\'enyi entropies with the insertion of a flux $\alpha$.
  The numerical results (symbols) for several values of $\alpha$ and $n$ are reported as function of $\ell$ for the filling $k_F=\pi/2$ (left) and $k_F=\pi/3$ (right). 
  The numerical data match well the Fisher-Hartwig prediction (cf. Eq. \eqref{tn_0}) although large oscillating corrections to the scaling are present. 
  }
  \label{Fig:snal}
\end{figure}

\subsubsection{Leading corrections $(m= \pm1)$}

The leading correction to the determinant $D_\ell(\lambda)$ comes from the terms with $m=\pm1$ in \eqref{gen_FHconj}
and is given by \cite{ce-10} 
\begin{multline}
\label{expa}
D_\ell(\lambda)
\simeq
D_\ell^{(0)}(\lambda)[1+ \Psi_\ell(\lambda)],\\
 \Psi_\ell(\lambda)=e^{-2i k_F \ell} L_k^{-2(1+2\beta_\l)} 
\frac{\Gamma^2(1+\beta_\l)}{\Gamma^2(-\beta_\l)}+
e^{2i k_F \ell} L_k^{-2(1-2\beta_\l)} 
\frac{\Gamma^2(1-\beta_\l)}{\Gamma^2(\beta_\l)}.
\end{multline}
We define the difference
\begin{align}
d_n (\alpha) \equiv \ln Z_n(\alpha)-\ln Z_n^{(0)}(\alpha),
 \end{align}
that for large $L_\kappa$ is
\bea
d_n (\alpha)\simeq
 \frac1{2\pi i}\oint d\lambda\ f_n(\l,\alpha) \frac{d\ln 
\left[1+ \Psi_\ell(\lambda)\right]}{d\l} 
=\frac1{2\pi i}\oint d\lambda\ f_n(\l,\alpha) \frac{d
\Psi_\ell(\lambda)}{d\l}+\cdots.
\label{approx}
\eea
\begin{figure}[tbp]
  \centering
  \includegraphics[width=\textwidth]{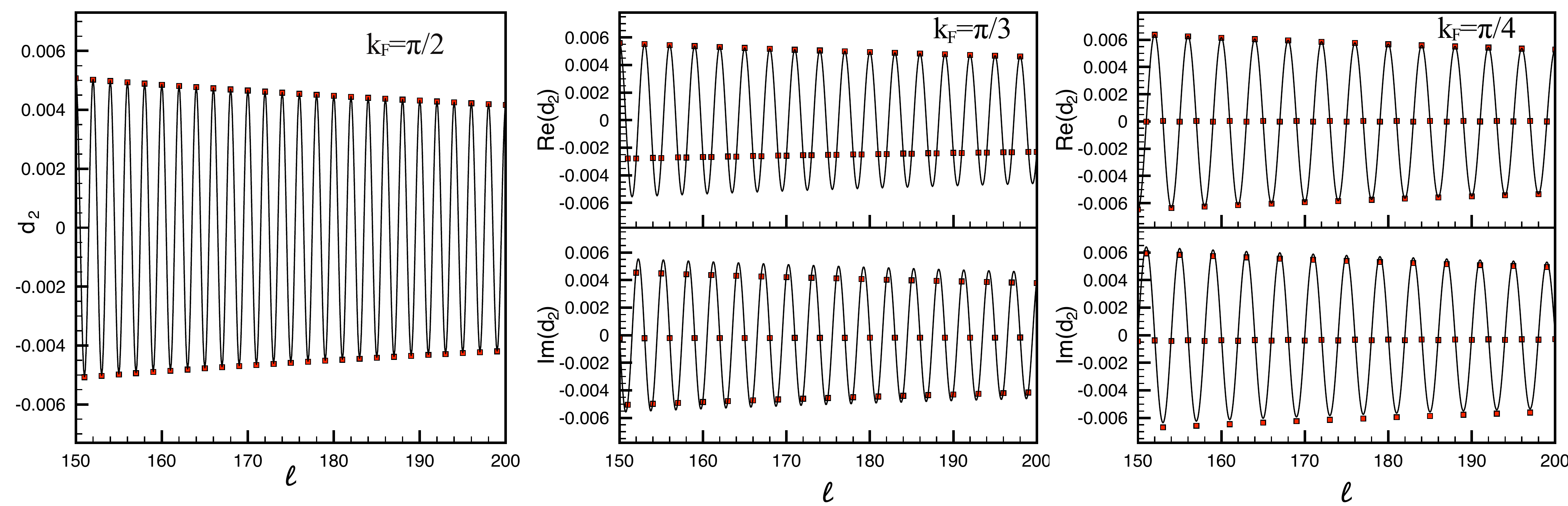}\\
  \caption{Behaviour of the leading corrections to the scaling. 
  The difference $d_2(\ell)\equiv \ln Z_2(\alpha)- \ln Z^{(0)}_2(\alpha)$ is reported for $\alpha =1$ and $k_F=\pi/2$ (left), $k_F=\pi/3$ (middle),
  $k_F=\pi/4$ (right) as a function of $\ell$. 
  The numerical data (symbols) perfectly match the calculated leading correction to the scaling from generalised Fisher-Hartwig 
  conjecture in Eq. \eqref{corrections} both for real and imaginary part.
  }
  \label{Fig:d2al1}
\end{figure}
Changing variable to $\lambda = \tanh (\pi w)$, in the final integral we only need the discontinuities across the branch cut that for the two cases are
\bea
\left[L_\kappa^{-2-4\beta} \frac{\Gamma^2(1+\beta)}{\Gamma^2(-\beta)}\right]_{\beta=-i w-\frac12}-
\left[L_\kappa^{-2-4\beta} \frac{\Gamma^2(1+\beta)}{\Gamma^2(-\beta)}\right]_{\beta=-i w+\frac12}&\simeq&
L_\kappa^{4 i w} \gamma^2(w),
\nonumber \\
\left[L_\kappa^{-2+4\beta} \frac{\Gamma^2(1-\beta)}{\Gamma^2(\beta)}\right]_{\beta=-iw-\frac12}-
\left[L_\kappa^{-2+4\beta} \frac{\Gamma^2(1-\beta)}{\Gamma^2(\beta)}\right]_{\beta=-iw +\frac12}&\simeq&
-L_\kappa^{-4 i w} \gamma^2(-w), 
\nonumber
\eea
where we have dropped terms of order $O(L_k^{-4})$ compared to the
leading ones and we have defined
\beq
\gamma(w)=\frac{\Gamma(\frac12-i w)}{\Gamma(\frac12+i w)}.
\eeq
Integrating by parts and using \eqref{Dwx} we finally get 
\begin{multline}
 d_n(\alpha) \simeq \frac{i n}{2}\int_{-\infty}^{\infty} \mbox{d}w\left( \tanh(\pi w)-\tanh(\pi n w+ i \alpha/2)\right) \\ \times
 \left[e^{-2 i k_F \ell}L_k^{4iw} \gamma^2(w)- e^{2i k_F \ell}L_k^{-4iw}\gamma^2(-w) \right].
\end{multline}
This integral can be evaluated on the complex plane by residue theorem.
For the first piece of the integral in square bracket, we should close the contour in the upper half plane,  while for the second piece in the lower half plane. 
In principle we should sum over all residues inside the integration contour, but if we are interested in the limit of large $L_k$, we can limit ourself to 
consider the singularities closest to the real axis. 
For the first integral this is at $w= i /(2n) (1-\alpha/\pi)$ while for the second one it is at $w=-i/(2n) (1+\alpha/\pi)$.
Summing up the two contributions we finally have 
\begin{equation}
\label{corrections}
d_n (\alpha)=  e^{- 2 i k_F \ell} L_k^{-\frac{2}{n}\left( 1-\frac{\alpha}{\pi} \right)}\left[ \frac{\Gamma\left( \frac{1}{2}+\frac{1}{2n}-\frac{\alpha}{2 \pi n}\right)}{\Gamma\left(  \frac{1}{2}-\frac{1}{2n}+\frac{\alpha}{2 \pi n}\right)}\right]^2 +
e^{2 i k_F \ell} L_k^{-\frac{2}{n}\left( 1+\frac{\alpha}{\pi} \right)}\left[ \frac{\Gamma\left( \frac{1}{2}+\frac{1}{2n}+\frac{\alpha}{2 \pi n}\right)}{\Gamma\left(  \frac{1}{2}-\frac{1}{2n}-\frac{\alpha}{2 \pi n}\right)}\right]^2.
\end{equation}
Let us comment this result. First, it is obvious that the formula is valid only for $-\pi<\alpha<\pi$, else one of the power laws would blow up as a consequence
of the fact that one of the terms for $m=\pm1$ becomes the leading ones.
Then we see that this correction is real only for $\alpha=0$ and at half-filling (when it should be real at all orders, cf. Eq. \eqref{sn_alpha2}). 
Away from half filling, there is generically a non-zero imaginary part.
The two contributions have a different power-law decays (for $\alpha\neq 0$) and so only one of them is the leading correction depending on the sign of $\alpha$.
However, for $\alpha$ close to zero, the two powers are too close in magnitude and they should be both taken into account in order to have an accurate 
description of the data for moderately large values of $\ell$. 
When $\alpha$ gets closer and closer to $\pm \pi$, Eq. \eqref{corrections} becomes accurate only for very large $\ell$ because the term with $m=0$ is about of the same 
order of magnitude as the one with $m=\pm1$ (depending on the sign of $\alpha$). A better description of the asymptotic behaviour may be achieved using 
Eq. \eqref{gen_FHconj} without expanding as in Eq. \eqref{expa}.
Finally, let us notice that, while in the absence of flux ($\alpha=0$) the oscillating corrections to the scaling vanish in the limit $n\to1$ \cite{ce-10},
for $\alpha\neq 0$ also the von Neumann entropy presents leading oscillating corrections described by Eq. \eqref{corrections}. 

\begin{figure}[tbp]
  \centering
  \includegraphics[width=\textwidth]{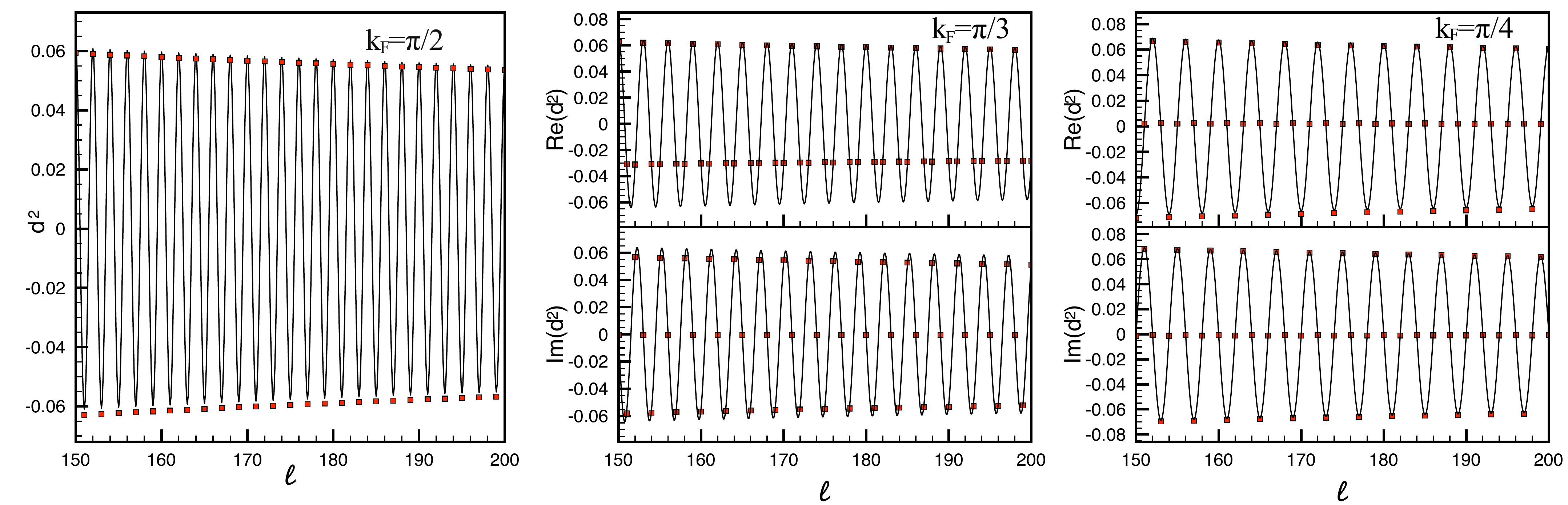}
  \caption{Behaviour of the leading corrections to the scaling. 
  The difference $d_2(\ell)\equiv \ln Z_2(\alpha)- \ln Z^{(0)}_2(\alpha)$ is reported for $\alpha =2$ and $k_F=\pi/2$ (left), $k_F=\pi/3$ (middle),
  $k_F=\pi/4$ (right) as a function of $\ell$. The numerical data (symbols) perfectly match the calculated leading correction to the scaling from generalised Fisher-Hartwig 
  conjecture in Eq. \eqref{corrections} both for real and imaginary part. We notice that here Eq. \eqref{corrections} works slightly worst compared to the case $\alpha=1$ in 
  Fig. \ref{Fig:d2al1}. 
  }
  \label{Fig:d2al2}
\end{figure}

In Figures \ref{Fig:d2al1} and \ref{Fig:d2al2} we report the difference $d_2(\alpha)$ as calculated numerically for $k_F=\pi/2,\pi/3, \pi/4$ and for 
$\alpha=1$ and $\alpha=2$ as function of $\ell$.
The numerical data are compared with the leading prediction in Eq. \eqref{corrections} and the agreement is extremely good. 
We actually observe that this prediction works slightly worst for $\alpha=2$ (cf. Fig. \ref{Fig:d2al2}) than for $\alpha=1$ (cf. Fig. \ref{Fig:d2al1}).
We indeed checked that the match becomes worst and worst when $\alpha$ moves close to $\pi$, when the leading term in the generalised Fisher-Hartwig changes. 
In principle it is possible to systematically analyse further corrections to $\ln Z_n(\alpha)$ by taking into account the known expansion of $D_\ell(\lambda)$ in powers 
of $\ell$ \cite{ce-10}, but this is very cumbersome and far beyond the scope of this paper. 

\section{Free fermions on a lattice: symmetry resolved entropies}
\label{sec:symm_resolved}

In this section we finally move to the symmetry resolved entropies and to their analysis. 

\subsection{$Q_A$-resolved moments via Fourier trasform}

The first step toward the symmetry resolved entropies is to calculate $\mathcal{Z}_n (q)$, the Fourier transform of $Z_n (\alpha)$ as defined in Eq. \eqref{fourier}.
We will show that we may obtain a very accurate prediction by keeping only the $m=0$ term in \eqref{gen_FHconj}, but with all non-universal pieces.
Within this approximation the Fourier transform $\mathcal{Z}_n (q)$ is
\beq
\label{fourier2}
\mathcal{Z}_n (q) \simeq \int_{-\pi}^{\pi} \frac{d \alpha}{2 \pi} e^{-i q \alpha} Z_n^{(0)}(\alpha) =  
L_k^{-\frac16 \left(n-\frac1n\right)} \int_{-\pi}^{\pi} \frac{d \alpha}{2 \pi} \, e^{-i \left( q - \frac{k_F}{\pi} \ell \right)  \alpha - b_n \alpha^2 } e^{ \Upsilon (n, \alpha)},
\eeq
where we defined the ``bare variance''
\beq
b_n= \frac{2 }{n} \frac{1}{4 \pi^2} \log L_k.
\eeq
As a first step we use Eq. \eqref{Upsexp} to rewrite $\mathcal{Z}_n (q)$ as
\begin{multline}
\label{fourier25}
\mathcal{Z}_n (q) \simeq 
e^{ \Upsilon (n)} L_k^{-\frac16 \left(n-\frac1n\right)} \int_{-\pi}^{\pi} \frac{d \alpha}{2 \pi} \, e^{-i \left( q - \frac{k_F}{\pi} \ell \right)  \alpha - (b_n -\gamma_2(n)) \alpha^2 } 
e^{ \epsilon (n, \alpha)}\\
=Z_n^{(0)}(0) \int_{-\pi}^{\pi} \frac{d \alpha}{2 \pi} \, e^{-i \left( q - \frac{k_F}{\pi} \ell \right)  \alpha - b_n^R \alpha^2 }  g_n(\alpha)  ,
\end{multline}
where $Z_n^{(0)}(0)=Z_n^{(0)}(\alpha=0)= e^{ \Upsilon (n)} L_k^{-\frac16 \left(n-\frac1n\right)}$,  we defined the ``renormalised variance''
\beq
b_n^R \equiv b_n -\gamma_2 (n)\,,
\eeq
and $g_n(\alpha)\equiv e^{\epsilon (n,\alpha)}$.
%
Up to this point, we only rewrote the starting  expression \eqref{fourier2}.
We now proceed by treating the integral, for large subsystem size $\ell$, by means of the saddle point approximation.
When $\ell \gg 1$, the large parameter in \eqref{fourier25} is $b_n$. Furthermore, we assume that $g_n(\alpha)=1$, because we have shown
in the previous section that the function $\epsilon(n,\alpha) \ll 1$, cf. Fig. \ref{Fig:ups}.
Within this approximation 
we finally get
\begin{multline}
\label{fourier4}
\mathcal{Z}_n (q) \simeq Z_n^{(0)}(0) \int_{-\infty}^{\infty} \frac{d \alpha}{2 \pi} \, e^{-i \left( q - \frac{k_F}{\pi} \ell \right)  \alpha - b_n^R \alpha^2 } =
Z_n (0) e^{- \frac{ \left( q-  \bar q \right)^2 }{4 b_n^R}}  \sqrt{\frac{1}{4\pi b_n^R}} = \\
Z_n (0) \sqrt{\frac{n \pi}{2  (\ln L_k -2\pi^2n \gamma_2(n) )}} e^{-\frac{n \pi^2 (q- \bar q)^2 }{2  (\ln L_k -2\pi^2 n\gamma_2(n))} }\,, 
\end{multline}
where we defined $\bar q\equiv \langle Q_A \rangle = \ell k_F / \pi$.  

Eq. \eqref{fourier4} is one of the main results of this paper. Let us discuss its features.
First, in the limit $\ell\to\infty$, we recover the CFT result  \eqref{Znq_FT} for $K=1$, but with  the correct normalisation of $Z_n (0)$.
Although this normalisation was not previously known rigorously (at least to the best of our knowledge), it could have been easily guessed  from 
the results in the absence of flux (i.e., $\alpha=0$ of Ref. \cite{JK}).
The mean of the gaussian term is $\bar q$ and it is not changed compared to the result \eqref{Znq_FT}. Consequently,  the main new insight from Eq. \eqref{fourier4} is the prediction for the constant term to add to $\ln L_k$
(or equivalently, the multiplicative scale for $L_k$ as in \cite{equi-sierra}). 
Although this non-universal constant is a correction to the leading behaviour (expanding for large $L_k$, it gives a term going like $(\ln L_k)^{-1}$), 
it is very important: the $(\ln L_k)^{-1}$ decay is so slow that it must be taken into account even for very large $L_k$ in order to quantitatively 
describe the data, as we shall see. 

\begin{figure}[t]
  \centering
  \includegraphics[width=0.325\textwidth]{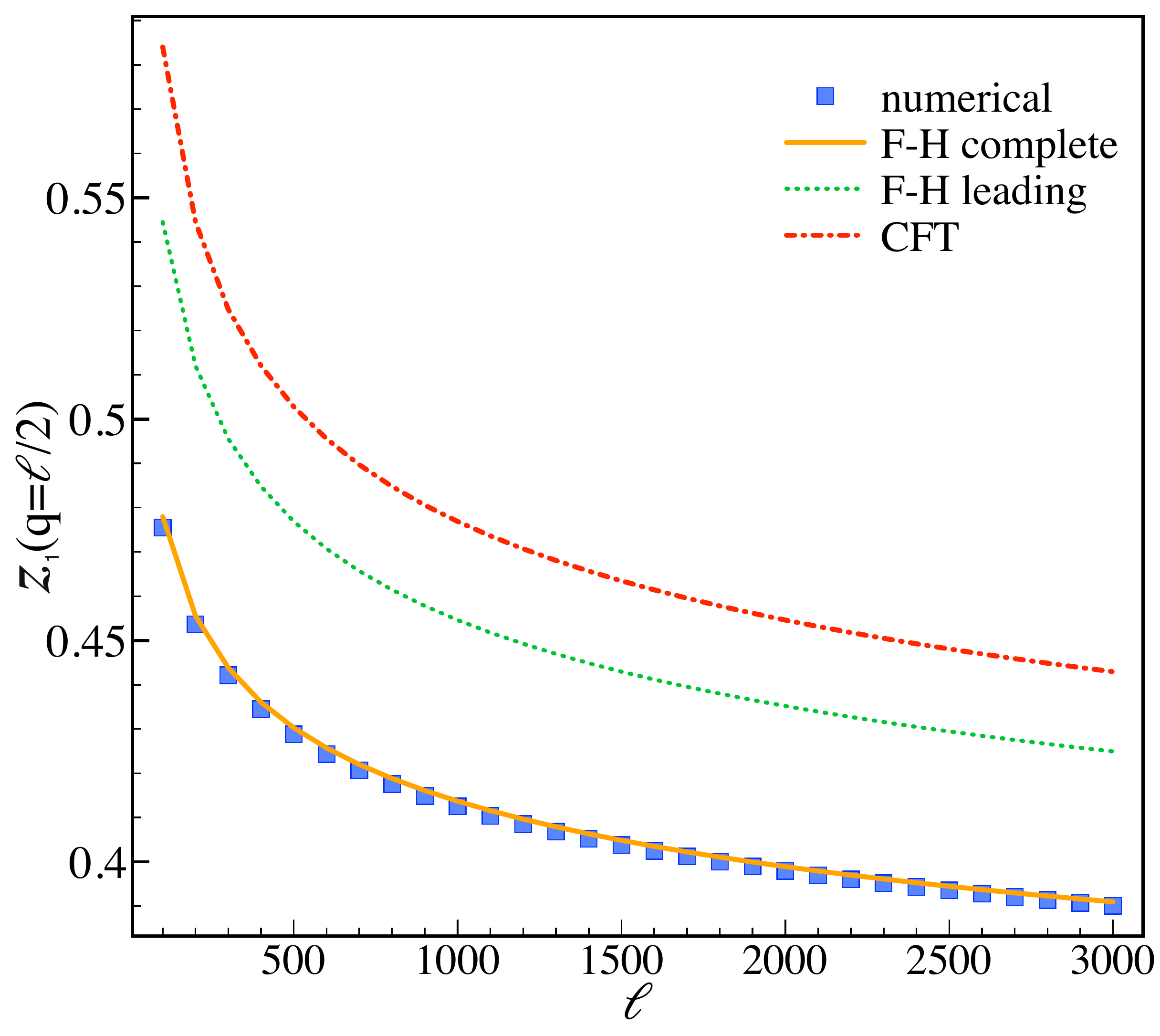}
   \includegraphics[width=0.325\textwidth]{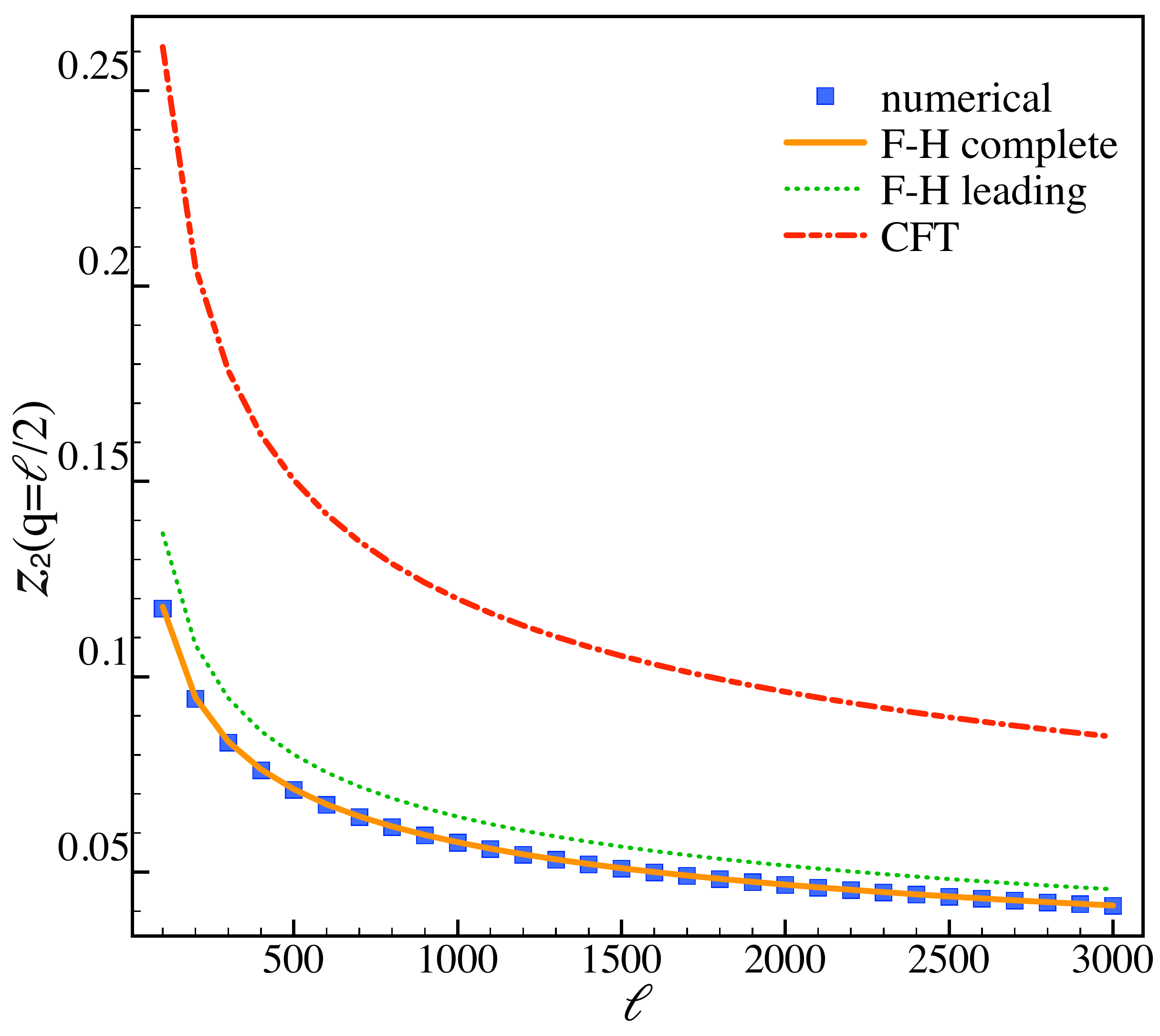}
    \includegraphics[width=0.325\textwidth]{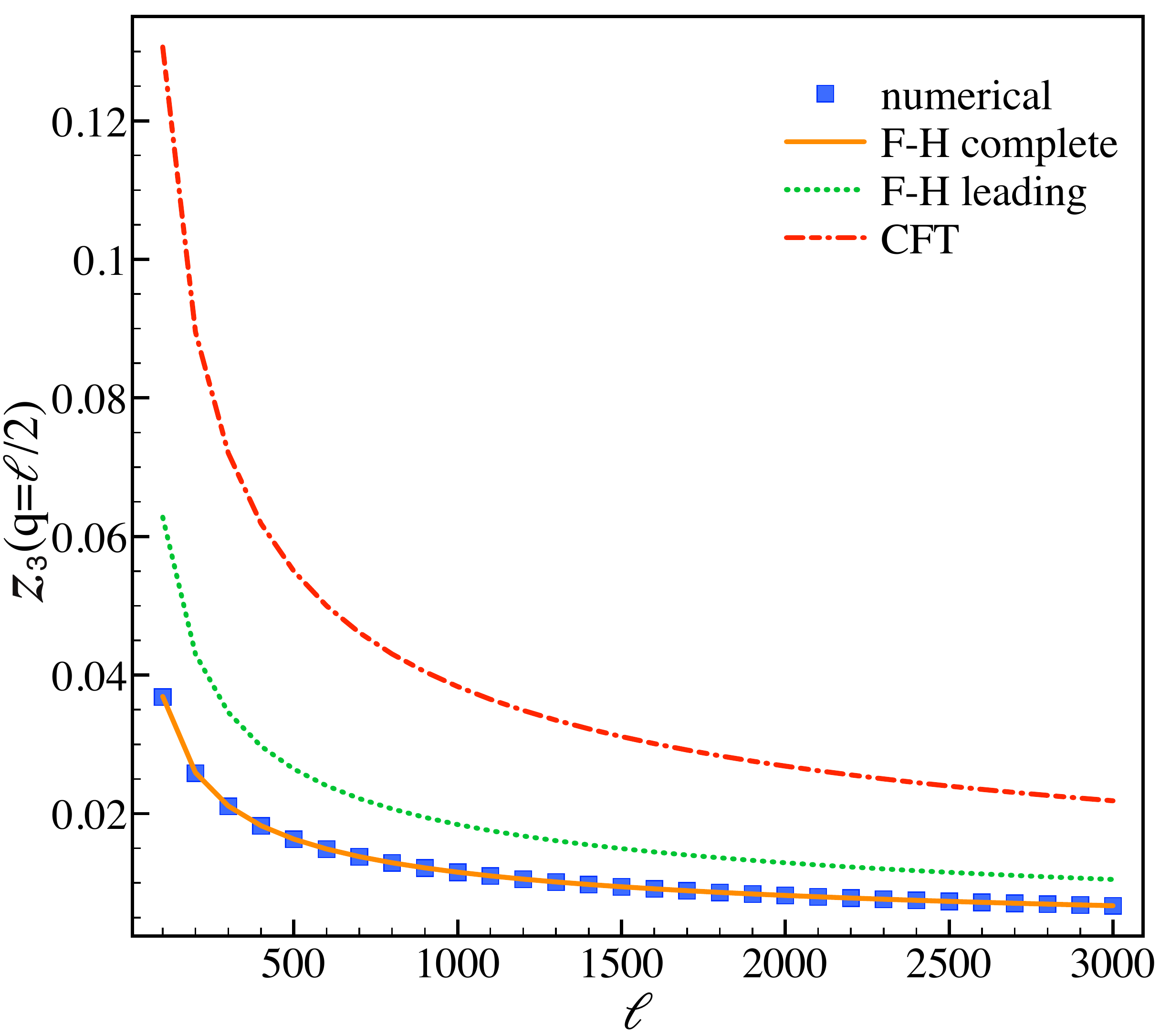}\\
     \includegraphics[width=0.325\textwidth]{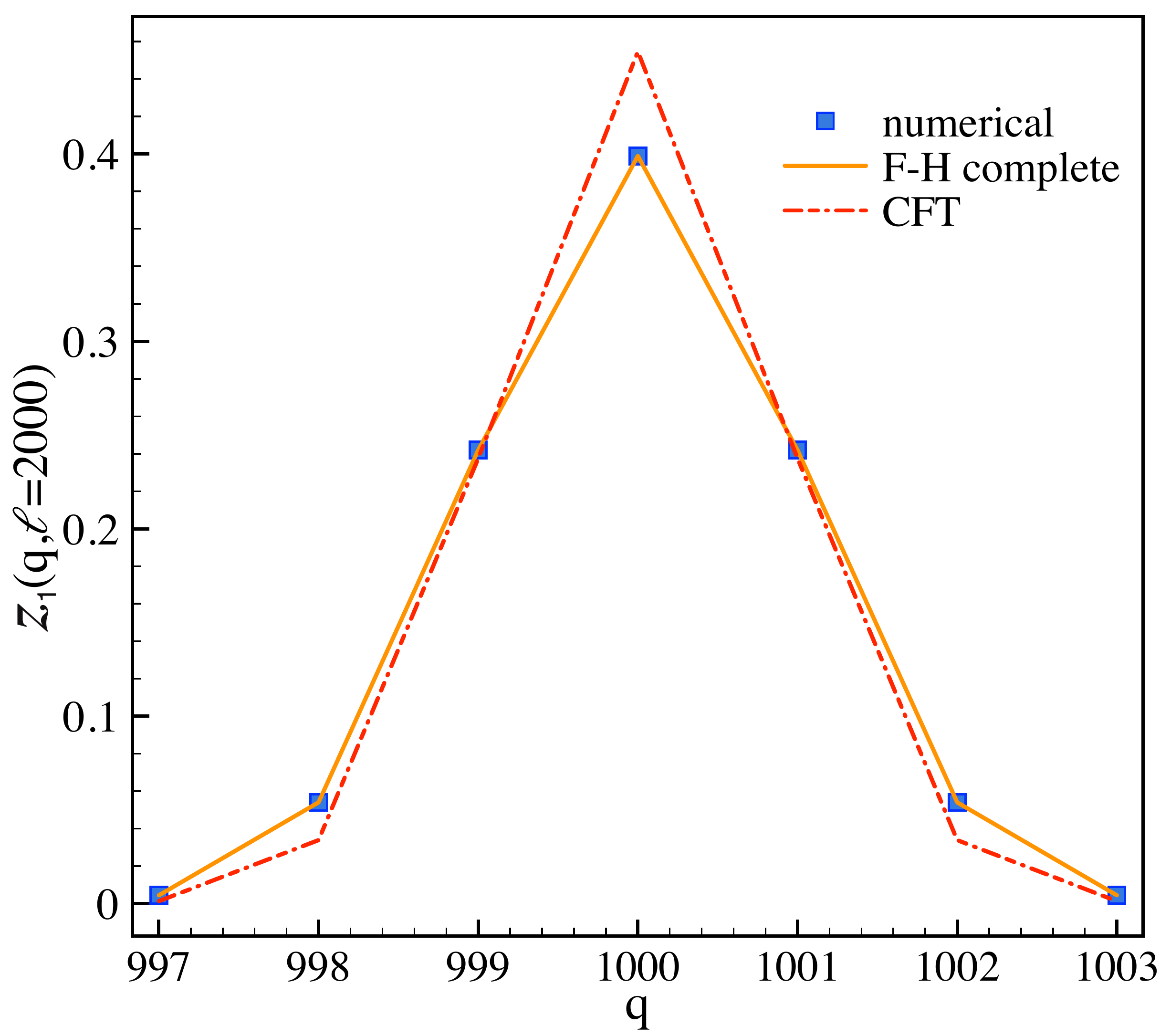}
   \includegraphics[width=0.325\textwidth]{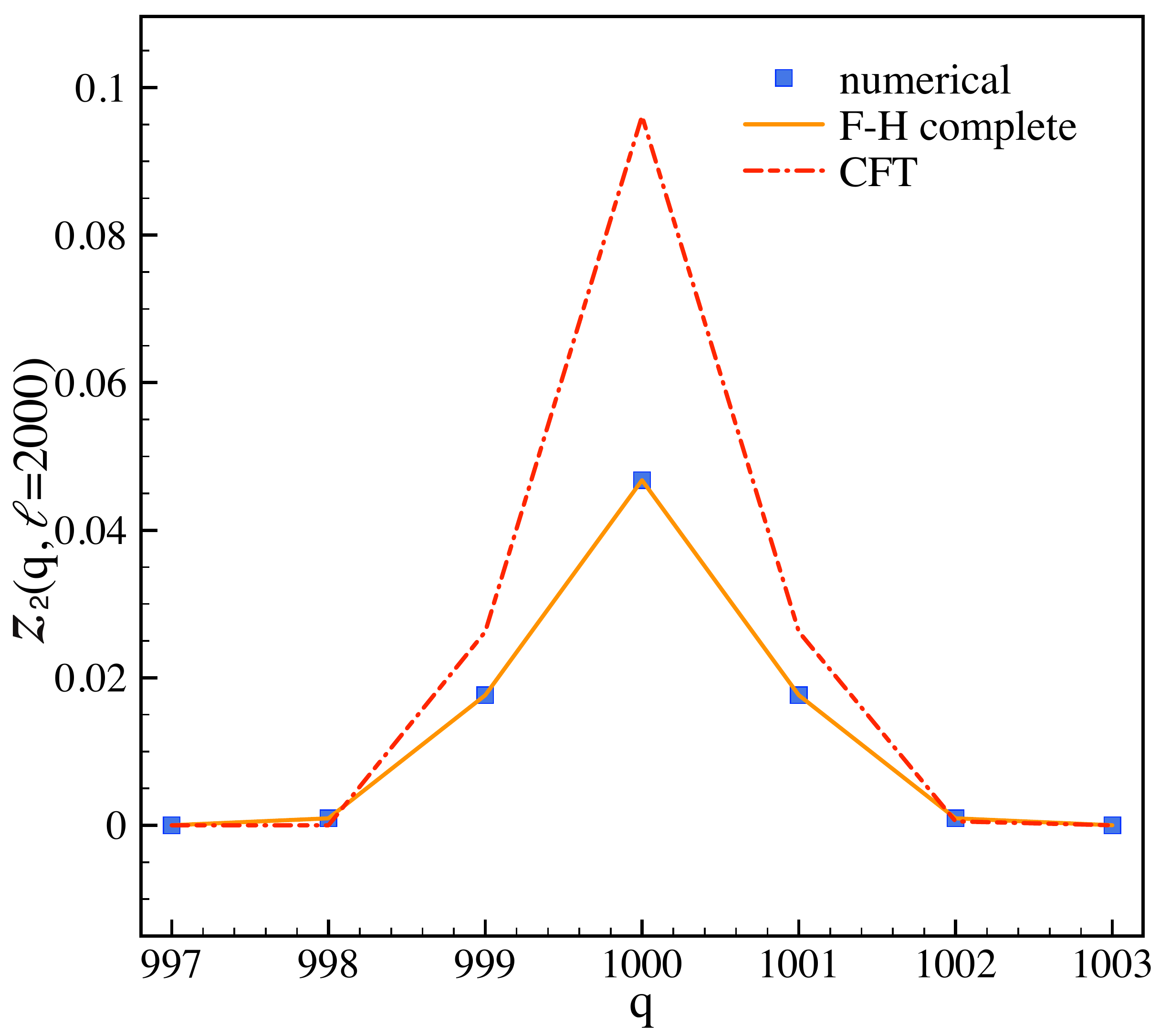}
    \includegraphics[width=0.325\textwidth]{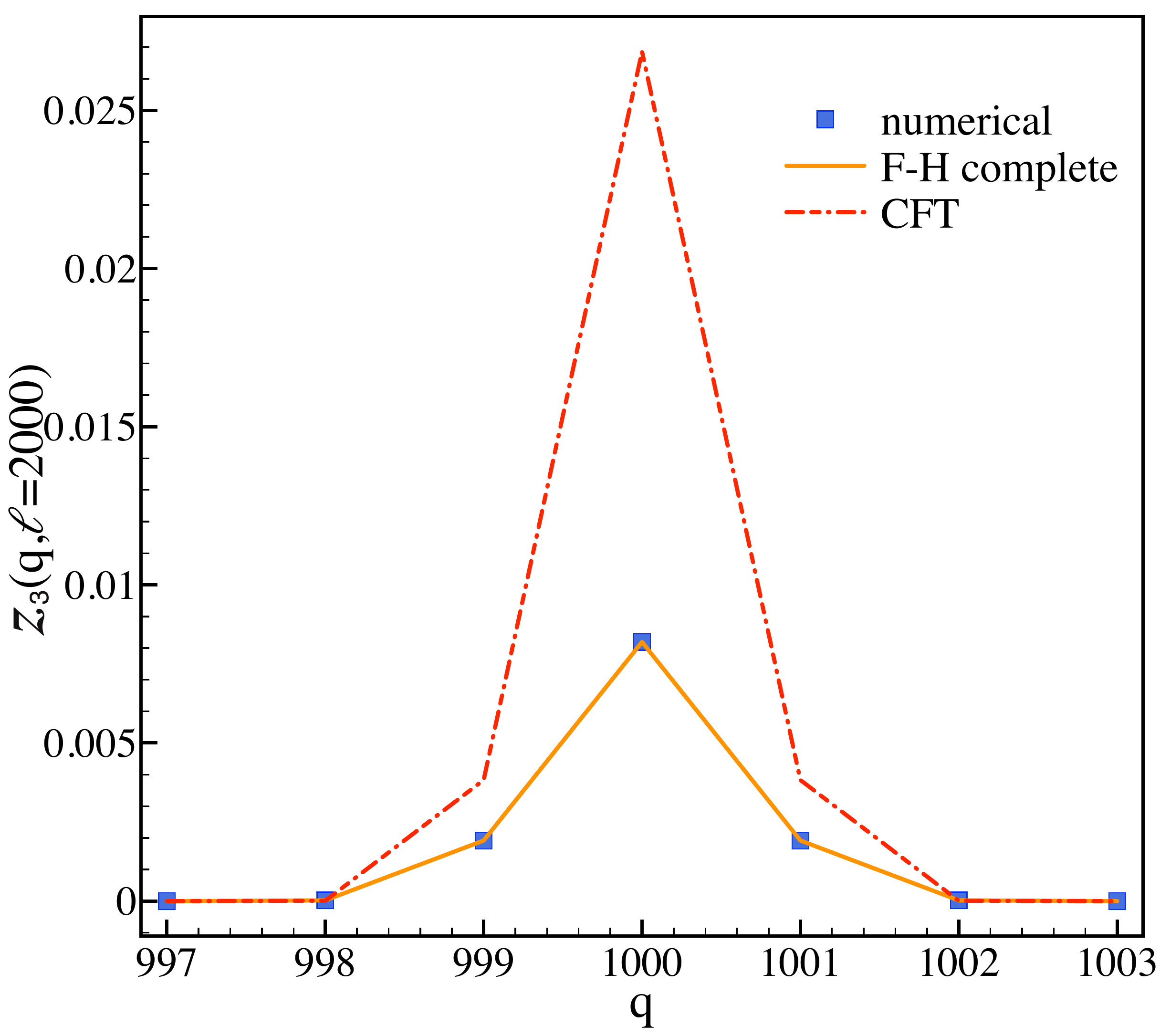}
  \caption{Symmetry resolved partition sums ${\cal Z}_n(q)$. Top: ${\cal Z}_n(q=\ell/2)$ at half filling as a function of $\ell$. 
  The numerical data for $n=1,2,3$ (left to right) are compared with: i) the CFT prediction without fixing the non universal constant (dot-dashed line);
  ii) the leading Fisher-Hartwig prediction at $O(1)$, i.e., Eq. \eqref{fourier4} with $\gamma_2(n)=0$ (dotted line);
  iii) the complete  Fisher-Hartwig result, Eq. \eqref{fourier4}  (full line). 
  Clearly only the latter accurately describes the data, although the qualitative behaviour is the same for all curves. 
  Bottom: ${\cal Z}_n(q)$ at half filling for $\ell=2000$ as function of $q$. 
  The numerical data for $n=1,2,3$ (left to right) are compared with: i) the CFT prediction without fixing the non universal constant (dot-dashed line);
  ii) the complete Fisher-Hartwig  result, Eq. \eqref{fourier4}  (full line).
  }
  \label{Fig:Zn}
\end{figure}

Given the importance that the quantity $\gamma_2(n)$ has in this analysis, we report its analytic expression
\begin{equation}
\label{gam2}
 \gamma_2{(n)}= \frac{n i}4\int_{-\infty}^\infty  dw [\tanh^3(\pi nw)-\tanh (\pi n w)]  \ln \frac{\Gamma(\frac12 +iw)}{\Gamma(\frac12 -iw)} \,,
\eeq
as well as its explicit numerical value for some $n$: $\gamma_2(1)=-0.0799027$, $\gamma_2(2)=-0.0462208\dots$, and $\gamma_2(3)=-0.0319926\dots$ 
(in particular $\gamma_2(1)=-(1+\gamma_E)/(2\pi^2)$ with $\gamma_E$ the Euler constant, as anticipated in \cite{equi-sierra}).
The importance of this constant in the description of the numerical data, was understood already in \cite{equi-sierra}, where the authors define
$g_n=2 e^{-2 \pi^2 n\gamma_2(n)}$ and provide the analytic results for $n=1$, as well as a numerical estimate for $n=2$, i.e., $g_2\sim 12.39$ which 
is very close to the exact value that we have found $g_2=12.4022\dots$.


Let us briefly discuss the terms that have been neglected in the derivation of Eq.~\eqref{fourier4}.
The most relevant one comes from  having approximated $g_n(\alpha)$ with $1$. 
By expanding this function in powers of $\alpha$, it is immediate to realise that the series coefficient $\alpha^{2k}$ (with $k\geq 2$) provides a correction to the leading term 
of the order $(\ln L_k)^{-k}$. Anyhow, these factors influence little the final result because the amplitude of the various terms is very small. 
Another correction comes from the extremes of integration that we pushed to $\pm \infty$ instead of $\pm \pi$. 
Although their effect can be taken into account as done in Ref. \cite{equi-sierra}, they provide  corrections which decay as $e^{- \pi^2 b_n^R}/b_n^R$, 
i.e., algebraically in $L_k$, and so negligible at this level. 
Also the corrections due to the terms with $m\neq 0$ in \eqref{gen_FHconj} decay as power laws in $L_k$ and can be safely neglected at this stage.

In Figure \ref{Fig:Zn} we report the numerically calculated symmetry resolved partition sums ${\cal Z}_n(q)$.
We compare the numerical data for $n=1,2,3$ with the CFT prediction without fixing the non universal constant as in Eq. \eqref{Znq_FT}.
The qualitative agreement is reasonable, but quantitatively far. 
We also report the prediction for ${\cal Z}_n(q)$ at order $O(\ell^0)$: the curves moves closer to the numerical data, but the match is still not perfect. 
Only when we use the complete Fisher-Hartwig prediction \eqref{fourier4} (with the correct value of $\gamma_2(n)$), the data are perfectly reproduced.
As anticipated, including the logarithmic corrections is fundamental to have an accurate description of the data.
Also the $q$-dependence of ${\cal Z}_n(q)$ is perfectly captured by \eqref{fourier4} as shown in the lower panels of Figure \ref{Fig:Zn}.

\subsection{Symmetry resolved R\'enyi and entanglement entropy}

We now use Eq.~\eqref{fourier4} to calculate the symmetry resolved R\`enyi and the Von Neumann entropies.
Let us start from the former. Eq. \eqref{replica} implies 
\beq
S_n (q)= \frac{1}{1-n} \ln \left[\frac{\mathcal{Z}_n(q)}{\mathcal{Z}_1 (q)^n} \right]\simeq \frac1{1-n}\ln \frac{Z_n(0)}{(Z_1(0))^n} 
\frac{e^{- \frac{ \left( q-  \bar q \right)^2 }{4 b_n^R}}}{e^{- \frac{n \left( q-  \bar q \right)^2 }{4 b_1^R}}}
\frac{(4\pi b_n^R)^{-1/2}}{(4\pi b_1^R)^{-n/2} } \,.
\label{Sn1}
\eeq
The first ratio in \eqref{Sn1} just gives the total R\'enyi entropy of order $n$, with the right additive constant (and indeed this is  true at all orders).
The other $q$-independent term is 
\begin{multline}
\frac{1}{1-n}\ln \frac{(4\pi b_n^R)^{-1/2}}{(4\pi b_1^R)^{-n/2} }=- \frac12\ln\frac2\pi+\frac{\ln n}{2(1-n)}+\frac{1}{1-n}
\ln\frac{(\ln L_k-2\pi^2\gamma_2(1))^{n/2}}{(\ln L_k-2\pi^2 n\gamma_2(n))^{1/2}}\\
=- \frac12\ln \Big(\frac2\pi\ln \delta_n L_k\Big) +\frac{\ln n}{2(1-n)}+ \cdots.
\end{multline}
The constant $\delta_n$ has been introduced to resum partially the subleading corrections to the scaling and it is given by 
\beq
\ln \delta_n=  -\frac{2\pi^2 n(\gamma_2(n)-\gamma_2(1))}{1-n}\,.
\label{deltan}
\eeq
The last term is the ratio of the two Gaussian factors which is the only one depending on $q$.
For this last contribution we have 
\beq
\frac{1}{1-n}\ln e^{\frac{n \left( q-  \bar q \right)^2 }{4 b_1^R}- \frac{ \left( q-  \bar q \right)^2 }{4 b_n^R}}=
 \left( q-  \bar q \right)^2 \pi^4 \frac{n}{1 - n}  (\gamma_2(1)-n\gamma_2(n)) \frac1{\ln^2 \kappa_n L_k} +\dots\,,
\eeq
where the constant
\beq
\ln \kappa_n=- \pi^2 (\gamma_2(1)+ n \gamma_2(n)),
\label{kappan}
\eeq
has been introduced, again, to resum partially the subleading corrections.

Putting together the three pieces we have 
\begin{multline}
S_n(q)= S_n
- \frac{1}{2} \ln \left( \frac{2}{\pi} \ln \delta_n L_k \right)  +\frac{\ln n}{2(1-n)} + 
\left( q-  \bar q \right)^2 \pi^4 \frac{n(\gamma_2(1)-n\gamma_2(n))}{1 - n}   \frac1{\ln^2 \kappa_n L_k} +\cdots\,.
\label{Snsr}
\end{multline}
This equation not only predicts the leading diverging behaviour for large $\ell$ which was already known from CFT \cite{GS,equi-sierra} (cf. Eq. \eqref{result_CFT}), but 
also the non-universal additive constant, as well as the some subleading corrections in $\ln L_k$.
The latter are not only  important to correctly describe the data, but are also the leading $q$-dependent contributions.
So while the leading and finite terms satisfy the equipartition of entanglement \cite{equi-sierra}, within our approach we are able to identify the leading term that 
breaks this equipartition.

Taking now the limit for $n\to 1$, we get the von Neumann entropy 
\begin{equation}
S_{\rm vN} (q) =  S_{\rm vN}
  - \frac{1}{2} \ln \left( \frac{2}{\pi} \ln \delta_1  L_k \right)  - \frac{1}{2} +
 \left( q-  \bar q \right)^2 \pi^4 \frac{(\gamma_2(1)+\gamma'_2(1))}{\ln^2 \kappa_1 L_k}  +\cdots ,
\label{Svnq}
\end{equation}
with $\ln \delta_1=2\pi^2 \gamma'_2(1)$ and $\gamma'_2(1)=0.0545724$.

\begin{figure}[tbp]
  \centering
  \includegraphics[width=.49\textwidth]{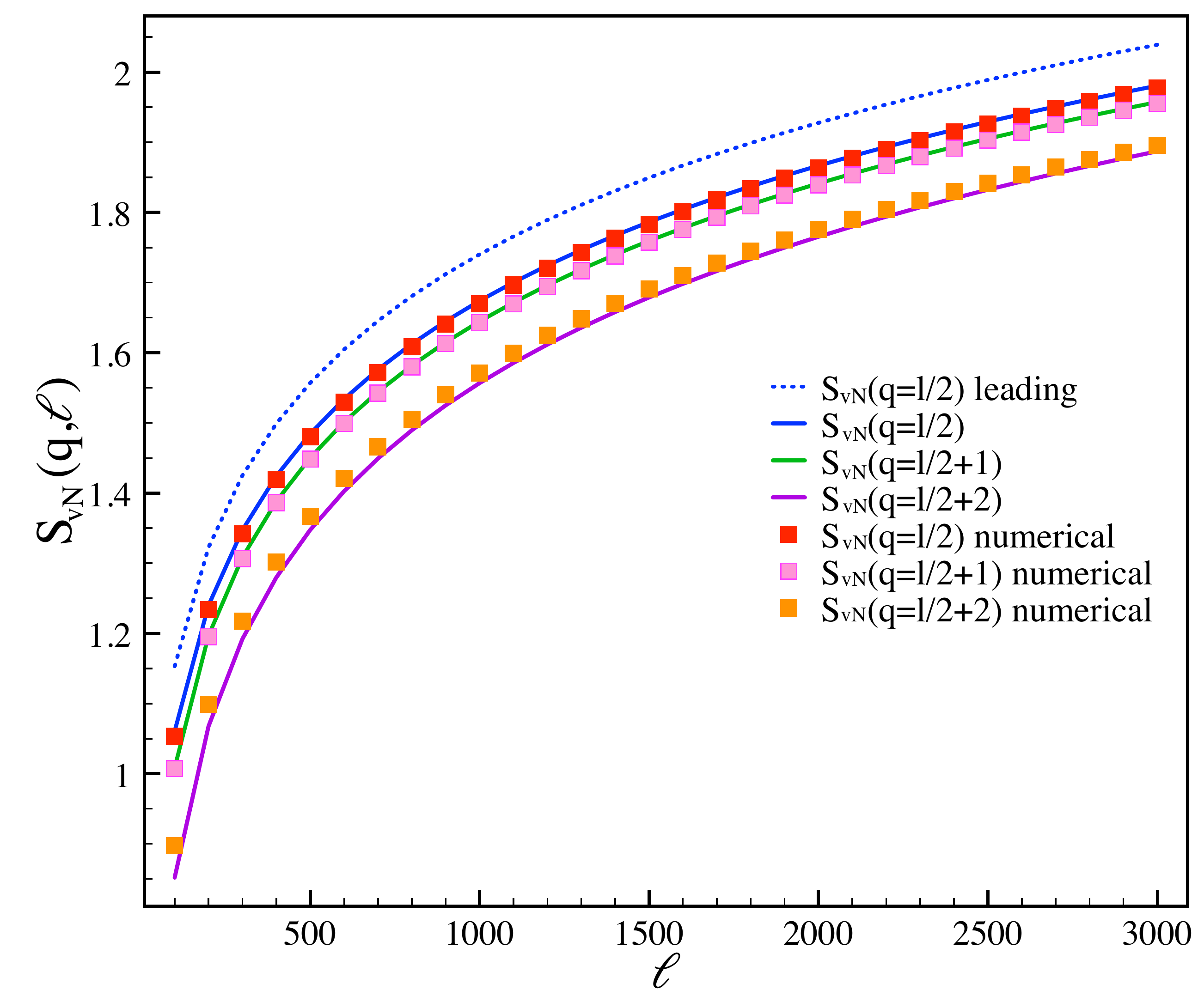}
  \includegraphics[width=.49\textwidth]{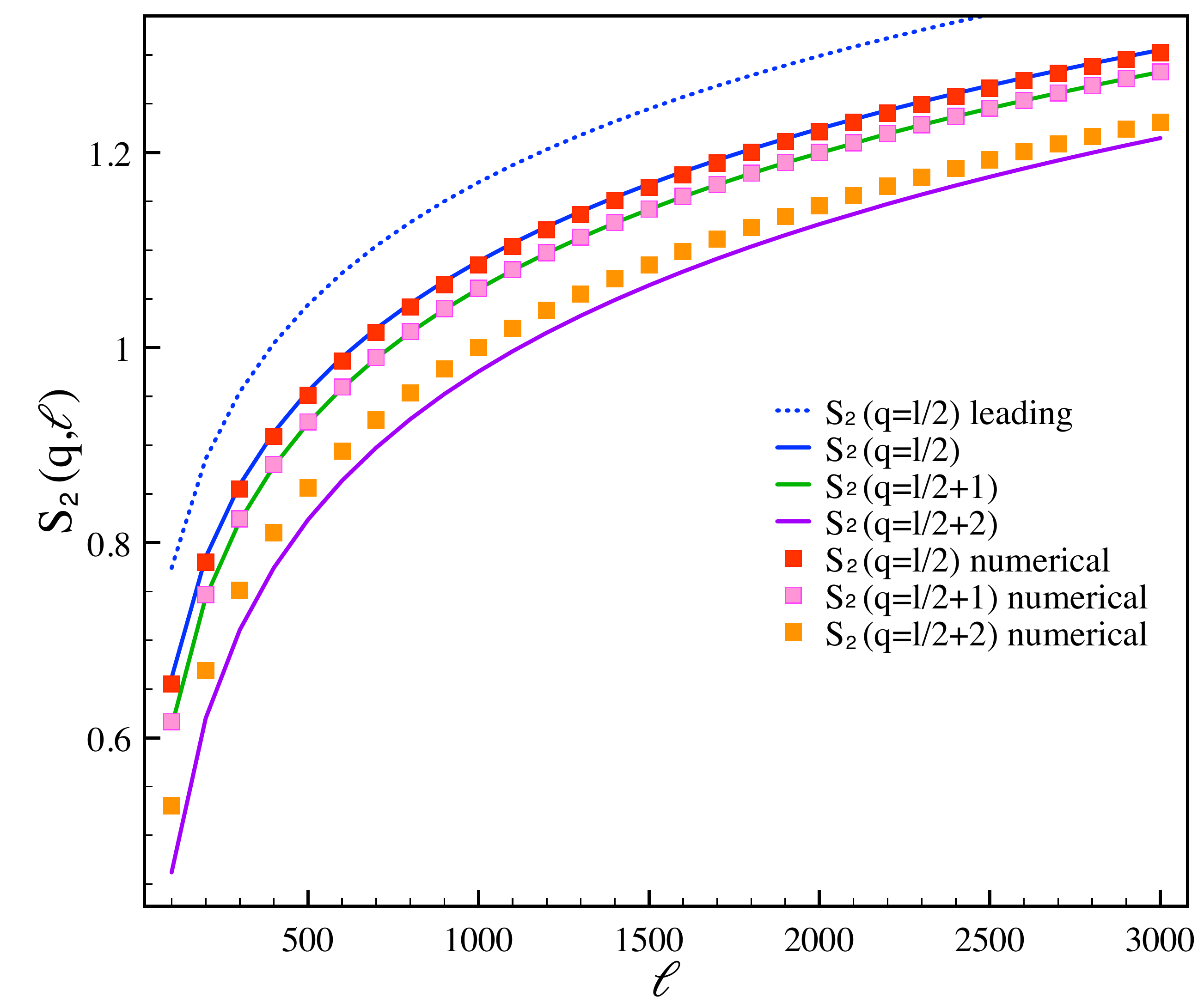}
  \caption{Von Neumann (left) and second R\'enyi (right) symmetry resolved entanglement entropies. 
  The numerical data (symbols) for $q=\ell/2, \ell/2+1, \ell/2+2$ are compared with the theoretical predictions 
  Eqs. \eqref{Snsr} and \eqref{Svnq}. The figures also highlight the importance of the logarithmic corrections to the 
  scaling which are fundamental in order to accurately describe the data for $\ell$ as large as 3000 and even larger.
  Increasing the values of $(q-\bar q)^2$, the corrections to the scaling that we neglect become more important. 
  }
  \label{Fig:Sq}
\end{figure}

These Fisher-Hartwig calculations for the symmetry resolved entanglement are compared with the numerical data in Figure  \ref{Fig:Sq}. 
It is evident in these figures that the results for different $q$ are {\it not} on top of each other although we reported $\ell$ as large as $3000$.
Indeed their difference (that we know to go to zero as $(\ln \ell)^{-2}$) can be easily misinterpreted as a different additive constant if one would proceed 
with a fit of the numerical data. Only the exact knowledge of the asymptotic behaviour \eqref{Snsr} and \eqref{Svnq} allow us to correctly understand the data.
In the figure we also report  Eqs. \eqref{Snsr} and \eqref{Svnq} truncated  at $o(1)$ (just for $q=\bar q$), showing that these leading curves are far from the data and that 
the distance between the two barely reduces. We stress that not only the prefactor of the logarithmic corrections are important, but also the precise values 
of the amplitudes  \eqref{kappan} and \eqref{deltan}, as it is easy to check. 
Finally we observe that increasing $(q-\bar q)$ the corrections to the scaling that we neglected become more important. 

We finish the section commenting about the double log contribution in Eqs. \eqref{Snsr} and \eqref{Svnq}.
It may seem rather awkward that all the symmetry resolved contributions have a double log correction, while the total entanglement entropy does not. 
Indeed, when calculating the total entanglement this double log cancels when summing to the fluctuation entanglement $S^f$ as in Eq. \eqref{decomposition}.
Indeed, taking as a prototypical example the von Neumann entropy and using that the probability is $p(q)={\cal Z}_1(q)$, we have
\begin{multline}
S^f= -\int_q {\cal Z}_1(q) \ln {\cal Z}_1(q)\simeq-\int dq \frac{e^{- \frac{ \left( q-  \bar q \right)^2 }{4 b_1^R}}}{\sqrt{4\pi b_1^R}} 
\ln  \frac{e^{- \frac{ \left( q-  \bar q \right)^2 }{4 b_1^R}}}{\sqrt{4\pi b_1^R}} 
=\frac12(1 +\ln  4\pi b_1^R)=\\
=\frac12\Big(1 +\ln \Big(\frac{2}{\pi}\ln L_k-\gamma_2(1)\Big)\Big)= \frac12 +\frac12\ln \Big(\frac{2}{\pi}\ln L_k\Big)+O(L_k^{-1})\,.
\end{multline}
Note that both leading terms in $S^f$ in the above equation cancel {\it exactly} with the corresponding ones in the symmetry resolved entanglement in Eq. \eqref{Svnq}.
The same is true for all R\'enyi entropies of arbitrary order.

\section{Charged and symmetry resolved entanglement for the Fermi gas}
\label{sec:gas}

In this section we derive the symmetry resolved entanglement entropy for a Fermi gas using the overlap matrix approach \cite{cmv-11l,cmv-11}.
This technique has been successfully applied to the calculation of entanglement in many different 
circumstances \cite{mrc-19,cmv-11l,cmv-11,cmv-11d,cmv-11c,v-12,ep-13,o-14,cw-16,lms-19}.

The system we are going to study consists of a gas of $N$ free spinless non-relativistic fermions with some suitable boundary conditions in order to have a discrete 
energy spectrum. 
The many body wave functions $\Psi(x_1,...,x_N)$ is the Vandermonde determinant $\Psi(x_1,...,x_N)={\rm det} [\phi_m(x_n)]/\sqrt{N!}$, 
built with the occupied single particle eigenstates with wave functions $\phi_m(x)$.
The many body ground state is obtained by filling the $N$ levels with lowest energies. 
Given that there is no lattice, the particle number $N$ provides also the ultraviolet cutoff.

Our focus is again a subsystem, now taken to be a single interval $A$, and its entanglement with the rest of the system.
The RDM $\rho_A$ in the subsystem $A$ is obtained as the continuous limit of Eq.~\eqref{RDM}. 
Therefore we can still exploit its gaussian nature.
A crucial quantity is the \emph{overlap matrix} associated to $A$ defined as
\beq
\mathbb{A}_{nm} = \int_A \phi^{*}_n (x) \phi_m (x)dx , \qquad n, m = 1, \cdots, N.
\eeq
In fact, it has been shown \cite{cmv-11,cmv-11l} that the (continuous) correlation matrix and the (discrete) overlap matrix share the same eigenvalues 
$(1+ \nu_k)/2 $ (with $k \in [1, N]$) and so, for example,  the moments of the RDM may be written as
\beq
\label{moments_gas}
{\rm tr} \rho_A^n = \prod_{i=1}^{N}  \left[ \left( \frac{1+\nu_i}{2}\right)^n  + \left( \frac{1- \nu_i}{2}\right)^n \right].
\eeq

In the case of a system with periodic boundary conditions in the interval $[0,L]$, the 
eigenfunctions are plane waves $\phi_{k}(x)=e^{2\pi i k x/L}/\sqrt{L}$ with integer wave-numbers $k$.
When the subsystem is also an interval, say $A=[0, \ell]$, the overlap matrix is easily calculated and reads
\beq
\mathbb{A}_{nm}= \frac{\sin \pi(n-m)\ell/L}{\pi (n-m)}, \qquad n, m = 1, \cdots, N.
\eeq
A crucial observation made in \cite{cmv-11} is that such matrix is identical to the lattice correlation matrix, Eq.~\eqref{cm_lattice}, upon identifying $k_F/ \pi$ with $\ell/L$ 
and $N$ with $\ell$.
As a consequence, this simple replacement allows to translate all the results from the lattice to the continuous model. 
In particular all formulas derived for the tight binding model are valid also for the gas, where now $L_k$ is not anymore $2\ell \sin k_F$, but 
\beq
L_k= 2N \sin \pi\frac{\ell}L\,.
\label{Lkgas}
\eeq
This replacement allows to show rigorously the CFT scaling in finite size for this model.

For the symmetry resolved entanglement we can straightforwardly make our predictions for the gas. 
First of all, the entanglement R\'enyi entropies in the presence of a flux are
\begin{multline}
\ln Z_n(\alpha)=  i \alpha \frac{ \ell N}{L}-
\left[ \frac{1}{6}\left( n-\frac{1}{n} \right) + \frac{2}{n}\left( \frac{\alpha}{2 \pi} \right)^2\right] \ln \Big[2N\sin\Big(\pi\frac\ell{L}\Big)\Big] +   \Upsilon{(n,\alpha)} +\\
e^{- \frac{2 i \pi \ell N}L } \Big[2N\sin\Big(\pi\frac\ell{L}\Big)\Big]^{-\frac{2}{n}\left( 1-\frac{\alpha}{\pi} \right)}\left[ \frac{\Gamma\left( \frac{1}{2}+\frac{1}{2n}-\frac{\alpha}{2 \pi n}\right)}{\Gamma\left(  \frac{1}{2}-\frac{1}{2n}+\frac{\alpha}{2 \pi n}\right)}\right]^2 +\\+
e^{\frac{2 i \pi \ell N}L} \Big[2N\sin\Big(\pi\frac\ell{L}\Big)\Big]^{-\frac{2}{n}\left( 1+\frac{\alpha}{\pi} \right)}\left[ \frac{\Gamma\left( \frac{1}{2}+\frac{1}{2n}+\frac{\alpha}{2 \pi n}\right)}{\Gamma\left(  \frac{1}{2}-\frac{1}{2n}-\frac{\alpha}{2 \pi n}\right)}\right]^2,
\end{multline}
where we only included the leading contributions at $m=0,\pm1$ in the generalised Fisher-Hartwig conjecture. 
We tested this result against exact numerical computations and, as for the lattice model, we found that it provides a very accurate description as long as $\alpha$ is not close 
to $\pm \pi$. 

Similarly, for the symmetry resolved entropies the predictions for the gas are obtained simply by plugging Eq. \eqref{Lkgas} into Eqs. \eqref{Snsr} and \eqref{Svnq}
for R\'enyi and von Neumann entropy respectively (with $\bar q= \frac{ \ell N}{L}$). 
%
%
%
%
%
%
%

\section{Conclusions}
\label{sec:end}

In this manuscript we derived exact formulas for the asymptotic behaviour of the symmetry resolved entanglement entropies in free fermion systems.
First we obtained an exact expression for the charged entropies given by Eqs. \eqref{tn_0} (asymptotic behaviour up to order $O(1)$) and \eqref{corrections} 
(leading corrections to the scaling). The leading logarithmic term in \eqref{tn_0} perfectly match the CFT prediction, but we also determined the non-universal $O(1)$ contribution. 
The $o(1)$ corrections present interesting oscillatory behaviour and a power law decay with exponents that depend on the flux $\alpha$.
We then moved to the truly symmetry resolved entropies  given by the Fourier transform of the charged ones.
The partition sums are given by Eq. \eqref{fourier4}, while R\'enyi anf von Neumann entropy by Eqs. \eqref{Snsr} and \eqref{Svnq} respectively.
These equations agree in the limit of large $\ell$ with the CFT results, but we also determine a number of non-universal constants as well as logarithmic corrections to 
the scaling which are fundamental for an accurate description of the numerical data. 
Our analysis also provide the first term in the expansion for large $\ell$ which depend on the symmetry sector,
hence breaking the equipartition of entanglement \cite{equi-sierra}. We also related the the double logarithmic correction to the fluctuation entanglement. 
 
While we have considered the specific case of free fermions, many features we find are in fact universal. 
The CFT results \cite{GS,equi-sierra} (cf. Eq. \eqref{result_CFT}) shows how the leading term of the charged entropies get renormalised by the Luttinger liquid parameter $K$.
A first natural question is how the exponent of the leading corrections to the scaling gets renormalised. 
It would be very interesting to adapt the field theoretical approach of Refs. \cite{cc-10,ot-15} to understand how this new universal exponent (equal to 
$2/n(1\pm\alpha/\pi)$ for free fermions) can be obtained in CFT.
For the symmetry resolved entanglement we showed the presence of very large logarithmic corrections to the scaling. 
The natrual question here is whether they are universal and if they can be also understood within CFT.
Furthermore, we find that many non-universal constants entering in these corrections are related to each other (e.g. Eqs. \eqref{deltan} and \eqref{kappan}): 
it is interesting to understand also the level of universality of these relations.

Finally there are few possible generalisations of the present calculations that can be done following the same logic as here; for example
the case of an open system can be analysed exploiting the generalised Fisher-Hartwig results in \cite{fc-11}, disjoint intervals using the approach in \cite{aef-14b}, and
trapped gases can be studied by random matrix techniques \cite{clm-15,lms-19} to recover results from curved CFT \cite{dsvc-17}.

\paragraph{Acknowledgments.}
We thank Luca Capizzi, Sara Murciano and Hassan Shapourian for useful discussions and collaboration on closely related topics. All authors acknowledge support from ERC under Consolidator grant number 771536 (NEMO).

\begin{appendix}

\section{Appendix A: Details of calculations}
\label{details}

In this appendix we report the details of the calculation needed in Section \ref{sec:flux}.

The first integral is the linear term in $\ell$
\beq
a_0= \frac{1}{2\pi i}\oint d\lambda f_n(\lambda,\alpha)\left(  \frac{1-k_F/\pi}{1+\lambda}-\frac{k_F/\pi}{1-\lambda}\right)\,.
\eeq
While the first piece is analytic in $\lambda$ (because $f_n(-1,\alpha)=0$), the second piece has a simple pole in $\lambda=1$, leading to 
\beq
a_0= -\frac{k_F/\pi}{2\pi i}\oint d\lambda \frac{f_n(\lambda,\alpha)}{1-\lambda}= i\alpha\frac{k_F}{\pi} \,.
\eeq

The second integral is 
\beq
a_1=\frac{2}{\pi^2} \oint d\lambda f_n(\lambda,\alpha) \frac{\beta_\lambda}{1-\lambda^2},
\eeq
The only non zero contribution to the contour integral comes from the discontinuity at the cut of $\beta_\lambda$, 
\beq
\beta_{x\pm i\epsilon}= -i w(x) \mp \frac12\,, \qquad {\rm with}\qquad 
w(x)=\frac1{2\pi} \ln \frac{1+x}{1-x}\,,
\label{Dwx}
\eeq
and hence we finally  have 
\beq
a_1=\frac2{\pi^2}\int_{-1}^1 d\lambda\ \frac{f_n(\lambda,\alpha)}{1-\lambda^2}.
\eeq
This integral may be considered as the final answer, but indeed it may be simply evaluated analytically.
First one performs the change of variable  
\beq
w= \frac1{2\pi} \ln \frac{1+\lambda}{1-\lambda},\qquad
\lambda= \tanh(\pi w)\ ,\quad -\infty<w<\infty,
\label{lam2w}
\eeq
to get
\beq
a_1=\frac2{\pi} \int_{-\infty}^\infty d w f_n(\tanh(\pi w),\alpha)\,. 
\eeq
At this point, integrate by part using 
\beq
\frac{d}{dw} f_n(\tanh(\pi w),\alpha)= \pi n [\tanh(\pi n w+i \alpha/2)-\tanh(\pi w)]\,,
\label{Dfn}
\eeq
to get
\beq
a_1=\frac2n \int_{-\infty}^\infty d w w (\tanh(\pi w)-\tanh(\pi n w+i \alpha/2))\,. 
\eeq
Consider now the difference 
\begin{multline}
a_1-a_1|_{\alpha=0}= \frac2n \int_{-\infty}^\infty d w w (\tanh(\pi n w)-\tanh(\pi n w+i \alpha/2))=\\
 \frac2{\pi^2 n} \int_{-\infty}^\infty d z z (\tanh(z)-\tanh(z+i \alpha/2))\,.
\end{multline} 
The derivative with respect to $\alpha$ of the last integral is 
\beq
 \frac{i}{\pi^2 n} \int_{-\infty}^\infty d z \frac{z}{\cosh(z+i \alpha/2)}= - \frac{\alpha}{\pi^2 n}  \,. 
\eeq
Integrating back and using the boundary condition that the difference for $\alpha=0$ is zero by definition, we finally have
\beq
a_1-a_1|_{\alpha=0}=-\frac{\alpha^2}{2\pi^2n}\,,
\eeq
that using the known value of $a_1|_{\alpha=0}$ \cite{JK} is equivalent to \eqref{tn_0}. 

The last integral is 
\beq
a_2=\frac{1}{\pi i}\oint  d\lambda f_n(\lambda,\alpha) \frac{d \ln [ G(1+\beta_{\lambda})G(1-\beta_{\lambda})]}{d\lambda}\,.
\eeq
Let us first integrate by parts 
\beq 
a_2=-\frac{1}{\pi i}\oint  d\lambda \frac{d f_n(\lambda,\alpha)}{d\lambda}   \ln [G(1+\beta_{\lambda})G(1-\beta_{\lambda})]\,.
\eeq
Again the only discontinuity at the cut comes from the function $\beta_\lambda$, so that 
\beq
\ln \frac{G(1+\beta_{x+i\epsilon})G(1-\beta_{x+i\epsilon})}{G(1+\beta_{x-i\epsilon})G(1-\beta_{x-i\epsilon})}=
\ln \frac{G(\frac12 -iw(x))G(\frac32 +iw(x))}{G(\frac32 -iw(x))G(\frac12 +iw(x))}=
\ln \frac{\Gamma(\frac12 +iw(x))}{\Gamma(\frac12 -iw(x))}\,,
\eeq
where we used $\Gamma(z)=G(z+1)/G(z)$.
We then have
\beq 
a_2=\frac{1}{\pi i}\int_{-1}^1  d\lambda \frac{d f_n(\lambda,\alpha)}{d\lambda}  \ln \frac{\Gamma(\frac12 +iw(x))}{\Gamma(\frac12 -iw(x))} \,.
\eeq
Changing integration variable from $\lambda$ to $w$ as in \eqref{lam2w} and using the derivative \eqref{Dfn}, we finally get
\beq
a_2={n i}\int_{-\infty}^\infty  dw [\tanh(\pi w)-\tanh (\pi n w+i\alpha/2)]  \ln \frac{\Gamma(\frac12 +iw)}{\Gamma(\frac12 -iw)} \,.
\eeq

\end{appendix}

%

%


\begin{thebibliography}{99}
%
\bibitem{amico-2008} 
L.~Amico, R.~Fazio, A.~Osterloh, and V.~Vedral, \emph{Entanglement in many-body systems},
\href{http://dx.doi.org/10.1103/RevModPhys.80.517}{Rev. Mod. Phys. {\bf 80}, 517 (2008).}

\bibitem{calabrese-2009} 
P.~Calabrese, J.~Cardy, and B.~Doyon Eds, \emph{Entanglement entropy in extended quantum systems}, 
\href{http://dx.doi.org/10.1088/1751-8121/42/50/500301}{J. Phys. A {\bf 42} 500301 (2009).}

\bibitem{eisert-2010}
J.~Eisert, M.~Cramer, and M.~B.~Plenio, \emph{Area laws for the entanglement entropy}, 
\href{http://dx.doi.org/10.1103/RevModPhys.82.277}{Rev. Mod. Phys. {\bf 82}, 277 (2010).}

\bibitem{rev-lafl}
N. Laflorencie, {\it Quantum entanglement in condensed matter systems}, 
\href{http://dx.doi.org/10.1016/j.physrep.2016.06.008}{Phys. Rep. {\bf 643}, 1 (2016)}.




\bibitem{greiner-15}
R. Islam, R. Ma, P. Preiss, M. Tai, A. Lukin, M. Rispoli, and M. Greiner, 
{\it Measuring entanglement entropy in a quantum many-body system},
\href{https://www.nature.com/articles/nature15750}{Nature { \bf 528}, 77 (2015).}

\bibitem{kauf-16}
A.~M. Kaufman, M.~E. Tai, A.~Lukin, M.~Rispoli, R.~Schittko, P.~M. Preiss,  and M. Greiner,
\textit{Quantum thermalization through entanglement in an isolated  many-body system},
  \href{http://dx.doi.org/10.1126/science.aaf6725}{Science {\bf 353} (2016) 794}, 
 

\bibitem{zoller-18}
A. Elben, B. Vermersch, M. Dalmonte, J. I. Cirac, and P. Zoller, 
{\it R\'enyi Entropies from Random Quenches in Atomic Hubbard and Spin Models}, 
\href{https://journals.aps.org/prl/abstract/10.1103/PhysRevLett.120.050406}{Phys. Rev. Lett. { \bf 120}, 050406 (2018)};\\
T. Brydges, A. Elben, P. Jurcevic, B. Vermersch, C. Maier, B. P. Lanyon, P. Zoller, R. Blatt, and C. F. Roos,
{ \it Probing R\'enyi entanglement entropy via randomized measurements},
\href{https://doi.org/10.1126/science.aau4963}{Science { \bf 364}, 6437 (2019)}.
%
\bibitem{exp-lukin}
A. Lukin, M. Rispoli, R. Schittko, M. E. Tai, A. M. Kaufman, S. Choi, V. Khemani, J. Leonard, and M. Greiner,
{\it Probing entanglement in a many-body localized system},
\href{https://science.sciencemag.org/content/364/6437/256/tab-figures-data}{Science, {\bf 364}, 6437 (2019)}.
%


\bibitem{cc-04}
P. Calabrese and J. Cardy, {\it Entanglement entropy and quantum field theory}, 
\href{http://dx.doi.org/10.1088/1742-5468/2004/06/P06002}{J.  Stat. Mech. P06002 (2004)}.

\bibitem{cc-09}
P. Calabrese and J. Cardy, {\it Entanglement entropy and conformal field theory}, 
\href{http://dx.doi.org/10.1088/1751-8113/42/50/504005}{J. Phys. A {\bf 42}, 504005 (2009)}.

\bibitem{hlw-94}
C. Holzhey, F. Larsen, and F. Wilczek, {\it Geometric and renormalized entropy in conformal field theory}, 
\href{http://dx.doi.org/10.1016/0550-3213(94)90402-2}{Nucl. Phys. B {\bf 424}, 443 (1994)}.

\bibitem{vlrk-03}
G. Vidal, J. I. Latorre, E. Rico, and A. Kitaev, {\it Entanglement in quantum critical phenomena}, 
\href{http://dx.doi.org/10.1103/PhysRevLett.90.227902}{Phys. Rev. Lett. {\bf 90}, 227902 (2003)};\\
J. I. Latorre, E. Rico, and G. Vidal,
{\it Ground state entanglement in quantum spin chains},
Quant. Inf. Comp. {\bf 4}, 048 (2004).



\bibitem{GS}
M. Goldstein and E. Sela, 
{\it Symmetry-Resolved Entanglement in Many-Body Systems}, 
\href{https://journals.aps.org/prl/abstract/10.1103/PhysRevLett.120.200602}{Phys. Rev. Lett. {\bf 120}, 200602 (2018)}.

\bibitem{CCAD08}
J. L. Cardy, O. A. Castro-Alvaredo, B. Doyon,
{\it Form Factors of Branch-Point Twist Fields in Quantum Integrable Models and Entanglement Entropy}, 
\href{https://doi.org/10.1007/s10955-007-9422-x}{ J Stat Phys (2008) 130: 129}

\bibitem{GS-neg}
M. Goldstein, E. Sela, 
{\it Imbalance Entanglement: Symmetry Decomposition of Negativity}, 
\href{https://doi.org/10.1103/PhysRevA.98.032302}{Phys. Rev. A {\bf 98}, 032302 (2018)}.

\bibitem{equi-sierra}
J. C. Xavier, F. C. Alcaraz, and G. Sierra, 
{\it Equipartition of the entanglement entropy}, 
\href{https://journals.aps.org/prb/abstract/10.1103/PhysRevB.98.041106}{Phys. Rev. B {\bf 98}, 041106 (2018)}.

\bibitem{fg-19}
N, Feldman and M. Goldstein,
{\it Dynamics of Charge-Resolved Entanglement after a Local Quench},
\href{https://arxiv.org/abs/1905.10749}{arXiv:1905.10749}.

\bibitem{CFH}
H. Casini, C. D. Fosco, and M. Huerta, 
{\it Entanglement and alpha entropies for a massive Dirac field in two dimensions}, 
\href{https://iopscience.iop.org/article/10.1088/1742-5468/2005/07/P07007}{J. Stat. Mech. (2005) P07007}.

\bibitem{ch-rev}
H Casini and M Huerta, {\it Entanglement entropy in free quantum field theory},
\href{https://doi.org/10.1088/1751-8113/42/50/504007}{J. Phys. A {\bf 42}, 504007 (2009)}.

\bibitem{d-16}
J. S. Dowker,  {\it Conformal weights of charged R\'enyi entropy twist operators for free scalar fields in arbitrary dimensions},
\href{https://doi.org/10.1088/1751-8113/49/14/145401}{J. Phys. A {\bf 49}, 145401 (2016)};\\
J. S. Dowker,  {\it Charged R\'enyi entropies for free scalar fields},
\href{https://doi.org/10.1088/1751-8121/aa6178}{J. Phys. A {\bf 50}, 165401 (2017)}.

\bibitem{matsuura}
A. Belin, L.-Y. Hung, A. Maloney, S. Matsuura, R. C. Myers, and T. Sierens, 
{\it Holographic charged R\'enyi entropies}, 
\href{https://doi.org/10.1007/JHEP12(2013)059}{JHEP {\bf 12} (2013) 059}.

\bibitem{cnn-16}
P. Caputa, M. Nozaki, and T. Numasawa, {\it Charged Entanglement Entropy of Local Operators},
\href{https://doi.org/10.1103/PhysRevD.93.105032}{Phys. Rev. D {\bf }93, 105032 (2016)}.



\bibitem{wv-03}
H. M. Wiseman and J. A. Vaccaro,
{\it Entanglement of Indistinguishable Particles Shared between Two Parties},
\href{https://journals.aps.org/prl/abstract/10.1103/PhysRevLett.91.097902}{Phys. Rev. Lett. {\bf 91}, 097902 (2003)}

\bibitem{ssr-17}
H. Shapourian, K. Shiozaki and S. Ryu, {\it Partial time-reversal transformation and entanglement negativity in fermionic systems}, 
\href{https://doi.org/10.1103/PhysRevB.95.165101}{Phys. Rev. B {\bf 95}, 165101 (2017)}.

\bibitem{shapourian-19}
H. Shapourian, P. Ruggiero, S. Ryu, and P. Calabrese,
{\it Twisted and untwisted negativity spectrum of free fermions},
\href{https://arxiv.org/pdf/1906.04211.pdf}{arXiv:1906.04211}

\bibitem{basor}
E. L. Basor and C. A. Tracy, 
{ \it The Fisher-Hartwig conjecture and generalizations}, 
\href{https://doi.org/10.1016/0378-4371(91)90149-7}{Physica A { \bf 177}, 167 (1991)};\\
E. L. Basor and K. E. Morrison, 
{ \it The Fisher-Hartwig conjecture and Toeplitz eigenvalues}, 
\href{https://doi.org/10.1016/0024-3795(94)90187-2}{Linear Algebra and Its Applications 202, 129 (1994)}.

\bibitem{JK}
B.-Q. Jin and V. E. Korepin, 
{ \it Quantum spin chain, Toeplitz determinants and Fisher-Hartwig conjecture}, 
\href{https://link.springer.com/article/10.1023/B:JOSS.0000037230.37166.42}{J. Stat. Phys. {\bf 116}, 79 (2004)}.

\bibitem{ce-10}
P. Calabrese and  F. H. L. Essler,
{\it Universal corrections to scaling for block entanglement in spin-1/2 XX chains},
\href{https://iopscience.iop.org/article/10.1088/1742-5468/2010/08/P08029}{J. Stat. Mech. (2010) P08029}.

\bibitem{lr-14}
N. Laflorencie and S. Rachel, {\it Spin-resolved entanglement spectroscopy of critical spin chains and Luttinger liquids},
\href{http://dx.doi.org/10.1088/1742-5468/2014/11/P11013}{J. Stat. Mech. (2014) P11013}.



\bibitem{peschel2001}
M. C. Chung and I. Peschel, \emph{Density-matrix spectra of solvable fermionic systems}, 
\href{http://dx.doi.org/10.1103/PhysRevB.64.064412}{Phys. Rev. {\bf B 64}, 064412 (2001).}

\bibitem{peschel2003}
I. Peschel, {\it Calculation of reduced density matrices from correlation functions},
\href{http://dx.doi.org/10.1088/0305-4470/36/14/101}{J. Phys. A {\bf 36}, L205 (2003)}.

\bibitem{pe-09}
I. Peschel and V. Eisler, {\it Reduced density matrices and entanglement entropy in free lattice models},
\href{http://dx.doi.org/10.1088/1751-8113/42/50/504003}{J. Phys. A {\bf 42}, 504003 (2009)};\\
I. Peschel, {\it Entanglement in solvable many-particle models},
\href{http://dx.doi.org/10.1007/s13538-012-0074-1}{Braz. J. Phys. 42, 267 (2012)}.

\bibitem{sfr-11a}
H. F. Song, C. Flindt, S. Rachel, I. Klich, and K. Le Hur,
{\it Entanglement from Charge Statistics: Exact Relations for Many-Body Systems},
\href{http://dx.doi.org/10.1103/PhysRevB.83.161408}{Phys. Rev. B {\bf 83}, 161408(R) (2011).}

\bibitem{sfr-11b}
H. F. Song, S. Rachel, C. Flindt, I. Klich, N. Laflorencie, and K. Le Hur,
{\it Bipartite Fluctuations as a Probe of Many-Body Entanglement},
\href{http://dx.doi.org/10.1103/PhysRevB.85.035409}{Phys. Rev. B {\bf 85}, 035409 (2012).}

\bibitem{cmv-12}
P. Calabrese, M. Mintchev and E. Vicari, {\it Exact relations between particle fluctuations and entanglement in Fermi gases},
\href{http://dx.doi.org/10.1209/0295-5075/98/20003}{EPL {\bf 98}, 20003 (2012).}

\bibitem{si-13}
R. Susstrunk and D. A. Ivanov,
{\it Free fermions on a line: Asymptotics of the entanglement entropy and entanglement spectrum from full counting statistics},
\href{https://doi.org/10.1209/0295-5075/100/60009}{EPL {\bf 100}, 60009 (2012).}

\bibitem{clm-15}
P. Calabrese, P. Le Doussal, and S. N. Majumdar, {\it Random matrices and entanglement entropy of trapped Fermi gases},
\href{https://doi.org/10.1103/PhysRevA.91.012303}{Phys. Rev. A {\bf 91}, 012303 (2015).}

\bibitem{km-05}
J. P. Keating and F. Mezzadri,
{\it Random Matrix Theory and Entanglement in Quantum Spin Chains},
\href{http://dx.doi.org/10.1007/s00220-004-1188-2}{Commun. Math. Phys. {\bf 252}, 543 (2004)};\\
J. P. Keating and F. Mezzadri,
{\it Entanglement in Quantum Spin Chains, Symmetry Classes of Random Matrices, and Conformal Field Theory},
\href{http://dx.doi.org/10.1103/PhysRevLett.94.050501}{Phys. Rev. Lett. {\bf 94}, 050501 (2005).}

\bibitem{afc-09}
V.~Alba, M.~Fagotti and P.~Calabrese, \textit{{Entanglement entropy of excited states}},
  \href{http://dx.doi.org/10.1088/1742-5468/2009/10/P10020}{J. Stat.  Mech. (2009) P10020}.

\bibitem{fc-11} M. Fagotti and P. Calabrese, 
{\it Universal parity effects in the entanglement entropy of XX chains with open boundary conditions}, 
\href{http://dx.doi.org/10.1088/1742-5468/2011/01/P01017}{J. Stat. Mech. P01017 (2011)}.

\bibitem{aef-14}
F. Ares, J. G. Esteve, F. Falceto, and E. Sanchez-Burillo, 
{\it Excited state entanglement in homogeneous fermionic chains},
\href{http://dx.doi.org/10.1088/1751-8113/47/24/245301 }{J. Phys. A {\bf 47} (2014) 245301}.

\bibitem{aef-14b}
F. Ares, J. G. Esteve, and F. Falceto, {\it Entanglement of several blocks in fermionic chains}, 
\href{http://dx.doi.org/10.1103/PhysRevA.90.062321 }{Phys. Rev. A {\bf 90} (2014) 062321}.

\bibitem{aef-15}
F. Ares, J. G. Esteve, F. Falceto, and A. R. de Queiroz,
{\it On the Mobius transformation in the entanglement entropy of fermionic chains},
\href{http://dx.doi.org/10.1088/1742-5468/2016/04/043106}{J. Stat. Mech. (2016) 043106};\\
F. Ares, J. G. Esteve, F. Falceto, and A. R. de Queiroz,
{\it Entanglement entropy and M\"obius transformations for critical fermionic chains},
\href{http://dx.doi.org/10.1088/1742-5468/aa71dc}{J. Stat. Mech. (2017) 063104}.

\bibitem{cl-08}
P. Calabrese and A. Lefevre, {\it Entanglement spectrum in one-dimensional systems},
\href{https://doi.org/10.1103/PhysRevA.78.032329}{Phys. Rev A {\bf 78}, 032329 (2008)}.


\bibitem{ccen-10}
P. Calabrese, M. Campostrini, F. Essler, and B. Nienhuis,
{\it Parity effect in the scaling of block entanglement in gapless spin chains}, 
\href{http://dx.doi.org/10.1103/PhysRevLett.104.095701}{Phys. Rev. Lett {\bf 104}, 095701 (2010)}.

\bibitem{cc-10}
J. Cardy and P. Calabrese, 
\emph{Unusual Corrections to Scaling in Entanglement Entropy}, 
\href{http://dx.doi.org/10.1088/1742-5468/2010/04/P04023}{J. Stat. Mech. (2010) P04023}.

\bibitem{ot-15}
K. Ohmori and Y. Tachikawa, {\it Physics at the entangling surface},
\href{http://dx.doi.org/10.1088/1742-5468/2015/04/P04010}{J. Stat. Mech P04010 (2015)}.

\bibitem{ccp-10} 
P. Calabrese, J. Cardy, and I. Peschel,
{\it Corrections to scaling for block entanglement in massive spin-chains},
\href{http://dx.doi.org/10.1088/1742-5468/2010/09/P09003}{J. Stat. Mech. (2010) P09003}.

\bibitem{cmv-11l}
P. Calabrese, M. Mintchev, and E. Vicari, {\it Entanglement Entropy of One-Dimensional Gases}, 
\href{https://doi.org/10.1103/PhysRevLett.107.020601}{Phys. Rev. Lett. {\bf 107}, 020601 (2011)}.

\bibitem{cmv-11}
P. Calabrese, M. Mintchev, and E.Vicari
{ \it The entanglement entropy of one-dimensional systems in continuous and homogeneous space},
\href{https://iopscience.iop.org/article/10.1088/1742-5468/2011/09/P09028}{J. Stat. Mech. (2011) P09028}.

\bibitem{cmv-11d}
P. Calabrese, M. Mintchev, and E. Vicari,
{\it Entanglement entropies in free fermion gases for arbitrary dimension}, 
\href{https://doi.org/10.1209/0295-5075/97/20009}{Europhys. Lett. {\bf 97} (2012) 20009}.


\bibitem{cmv-11c}
P. Calabrese, M. Mintchev, and E. Vicari,
{\it Entanglement entropy of quantum wire junctions},  
\href{https://doi.org/10.1088/1751-8113/45/10/105206}{J. Phys. A {\bf 45} (2012) 105206}.

\bibitem{v-12}
E. Vicari,
{\it Entanglement and particle correlations of Fermi gases in harmonic traps},
\href{https://doi.org/10.1103/PhysRevA.85.062104}{Phys. Rev. A {\bf 85}, 062104 (2012)}.

\bibitem{ep-13}
V. Eisler and I. Peschel,
{\it Free-fermion entanglement and spheroidal functions},
\href{https://doi.org/10.1088/1742-5468/2013/04/P04028}{J. Stat. Mech. (2013) P04028}.

\bibitem{o-14}
A. Ossipov, {\it Entanglement Entropy in Fermi Gases and Anderson?s Orthogonality Catastrophe}
\href{https://doi.org/10.1103/PhysRevLett.113.130402}{Phys. Rev. Lett. 113, 130402 (2014)}.

\bibitem{cw-16}
P.-Y. Chang and X. Wen, {\it Entanglement negativity in free-fermion systems: An overlap matrix approach},
\href{https://doi.org/10.1103/PhysRevB.93.195140}{Phys. Rev. B {\bf 93}, 195140 (2016)}.

\bibitem{lms-19}
B. Lacroix-A-Chez-Toine, S. N. Majumdar, and G. Schehr,
{\it Entanglement Entropy and Full Counting Statistics for 2d-Rotating Trapped Fermions},
\href{https://doi.org/10.1103/PhysRevA.99.021602}{Phys. Rev. A {\bf 99}, 021602 (2019)}.

\bibitem{mrc-19}
S.~Murciano, P.~Ruggiero and P.~Calabrese, \textit{{Entanglement and relative
  entropies for low-lying excited states in inhomogeneous one-dimensional  quantum systems}},  
 \href{http://dx.doi.org/10.1088/1742-5468/ab00ec}{J. Stat. Mech. (2019) 034001}.

\bibitem{dsvc-17} J. Dubail, J.-M. St\'ephan, J. Viti, and P. Calabrese, 
{\it Conformal field theory for inhomogeneous one-dimensional quantum systems: the example of non-interacting Fermi gases}, 
\href{http://dx.doi.org/10.21468/SciPostPhys.2.1.002}{Scipost Phys. {\bf 2}, 002 (2017)}.


\end{thebibliography}
\end{document}